\documentclass[acmsmall, authorversion, screen]{acmart}

\usepackage{framed}
\usepackage{listings}
\usepackage{subcaption}

\usepackage[utf8]{inputenc}
\usepackage[T1]{fontenc}
\usepackage{microtype}

\usepackage{lipsum}
\usepackage{colortbl}
\usepackage{collcell}

\usepackage{pgfplots}
\pgfplotsset{compat=newest}

\usepackage[ruled,vlined,linesnumbered]{algorithm2e}
\usepackage{amsfonts}
\usepackage{amsthm}
\usepackage{natbib}
\usepackage{mathtools}
\usepackage{stmaryrd}
\usepackage{braket}
\usepackage{mathpartir}
\usepackage{bm}
\usepackage{IEEEtrantools}
\usepackage{hyperref}
\usepackage[nameinlink]{cleveref}
\usepackage{xcolor}
\usepackage{xparse}
\usepackage{etoolbox}
\usepackage{listings}
\usepackage{tabularx}
\usepackage{xfrac}
\usepackage{booktabs}
\usepackage{xspace}
\usepackage{tikz-cd}
\usepackage{pgf}
\usepackage{tikz}
\usepackage{wrapfig}
\usepackage{subcaption}
\usepackage{pifont}
\usepackage{hhline}
\usepackage{multirow}
\usepackage{enumitem}
\usepackage{ragged2e}
\usepackage{fancyvrb}

\setcopyright{cc}
\setcctype{by}
\acmDOI{10.1145/3720474}
\acmYear{2025}
\acmJournal{PACMPL}
\acmVolume{9}
\acmNumber{OOPSLA1}
\acmArticle{118}
\acmMonth{4}
\received{2024-10-16}
\received[accepted]{2025-02-18}

\makeatletter
\let\@authorsaddresses\@empty
\makeatother

\definecolor{DOrange}{RGB}{145,20,20}
\definecolor{DPurple}{RGB}{109,51,125}

\definecolor{CBPurple}{HTML}{702A5A}
\definecolor{CBRed}{HTML}{EE442F}
\definecolor{CBBlue}{HTML}{539BBE}

\newcommand{\kestrel}{\textsc{KestRel}}
\newcommand{\coreRel}{\textsc{CoreRel}}
\newcommand{\imp}{\textsc{Imp}\xspace}

\newbool{nofancystyle}
\newcommand{\fancystyle}[1]{\ifbool{nofancystyle}{}{#1}}

\newcommand{\cmstyle}{\booltrue{nofancystyle}\color{olive}}

\newcommand{\kwstyle}{\fancystyle{\color{violet}}}
\newcommand{\dfstyle}{\fancystyle{\color{magenta}}}
\newcommand{\symbolstyle}{\fancystyle{\color{darkgray}}}
\newcommand{\typestyle}{\fancystyle{\color{teal}}}
\newcommand{\metastyle}{\fancystyle{\color{blue}}}

\newcommand{\oblivstyle}[1]{$\widehat{\mbox{#1}}$}

\newcommand{\mutestyle}{\fancystyle{\color{gray}}}

\newcommand{\Ddownarrow}{\mathbin{\rotatebox[origin=c]{90}{$\Lleftarrow$}}}

\newcolumntype{C}{>{$}c<{$}}
\newcolumntype{L}{>{$}l<{$}}
\newcolumntype{R}{>{$}r<{$}}
\newcolumntype{P}[1]{>{\RaggedRight\arraybackslash}p{#1}}

\fvset{fontsize=\footnotesize}
\definecolor{lightergray}{gray}{0.9}

\newcommand{\production}{\Coloneqq}

\expandafter\patchcmd\csname \string\lstinline\endcsname{%
  \leavevmode
  \bgroup
}{%
  \leavevmode
  \ifmmode\hbox\fi
  \bgroup
}{}{%
  \typeout{Patching of \string\lstinline\space failed!}%
}

  \SetKwInOut{Params}{Inputs}
  \SetKwInOut{Output}{Output}
  \SetKwFunction{CreateEGraph}{CreateEGraph}
  \SetKwFunction{Insert}{Insert}
  \SetKwFunction{EQSat}{EQSat}
  \SetKwFunction{ExtractLocal}{ExtractLocal}
  \SetKwFunction{RandomStates}{RandomStartStates}
  \SetKwFunction{Cost}{Cost}
  \SetKwFunction{Temperature}{Temperature}
  \SetKwFunction{Neighbor}{RandomNeighbor}
  \SetKwFunction{Evaluate}{Evaluate}
  \SetKwFunction{Jump}{Jump}
  \SetKwFunction{Reify}{Reify}
  \SetKwFunction{Verify}{Verify}

  \SetKwFunction{Daikon}{Daikon}
  \SetKwFunction{Hints}{BaseInv}
  \SetKwFunction{Dafny}{Dafny}
  \SetKwFunction{Annotate}{Annotate}

\newcommand{\relBegin}[2]{\ensuremath \, \textsf{rB}\langle\lstinline!#1!, \lstinline!#2!\rangle}
\newcommand{\relMiddle}[2]{\ensuremath \, \textsf{rM}\langle\lstinline!#1!, \lstinline!#2!\rangle}
\newcommand{\relEnd}[2]{\ensuremath \, \textsf{rE}\langle \lstinline!#1!, \lstinline!#2!\rangle}

\newcommand{\whileBegin}{\textsf{wB}}
\newcommand{\whileHead}{\textsf{wH}}
\newcommand{\whileEnd}{\textsf{wE}}

\makeatother

\newcommand{\hltrip}[3]{\ensuremath{\vdash \left\{\lstinline[mathescape]{#1}\right\}\ \lstinline[mathescape]{#2}\ \{\lstinline[mathescape]{#3}\}}}
\newcommand{\relcomp}[0]{\ensuremath\mathbb{\textcolor{DPurple}{\bf \,\raisebox{-4pt}{$\stackrel{\bullet}{\textsf{\ding{124}}}$} \,}}}

\newcommand{\rulesep}{\unskip\ \vrule height -1ex\ }

\newcommand{\fail}[1]{\textcolor{red}{\ding{54} #1}}
\newcommand{\timeout}{\textcolor{red}{\ding{54} $\infty$}}
\definecolor{DarkGreen}{HTML}{006400}
\newcommand{\yay}[1]{\textcolor{DarkGreen}{\ding{52} #1}}

\newcommand{\NoOK}{\textcolor{red}{\ding{54}}}
\newcommand{\OK}{\textcolor{DarkGreen}{\ding{52}}}

\definecolor{darkblue}{HTML}{06038D}

\def\lstinlines|#1|{\text{\lstinline[basicstyle=\footnotesize]|#1|}}

\newlength{\relbracketwidth}
\newenvironment{relbrackets}[3][1]
{\def\relbracketscale{#1}
 \setlength{\relbracketwidth}{#2\linewidth}
 \addtolength{\relbracketwidth}{#3\linewidth}
 \begin{tabular}{m{.1cm}m{\relbracketwidth}m{.1cm}}
   \scalebox{1}[\relbracketscale]{\textcolor{DPurple}{$\langle$}}
   & \vspace{-.4cm}
     \hspace{-.4cm}
     \begin{minipage}{\relbracketwidth}
    \begin{tabular}{p{#2\linewidth}|p{#3\linewidth}}}
{\end{tabular}
\end{minipage}
   & \hspace{-.2cm}
     \scalebox{1}[\relbracketscale]{\textcolor{DPurple}{$\rangle$}}
 \end{tabular}
}

\newenvironment{ssinglebracketsl}[2][\linewidth]
{\def\relbracketscale{#2}
 \begin{tabular}{m{.1cm}m{#1}m{.1cm}}
   \scalebox{1}[\relbracketscale]{\textcolor{DPurple}{$\langle$}}
   & \vspace{-.4cm}
     \hspace{-.4cm}
  \begin{minipage}{\linewidth}
    \begin{tabular}{p{\linewidth}}}
{\end{tabular}
\end{minipage}
   & \hspace{-.2cm}
     \scalebox{1}[\relbracketscale]{\textcolor{DPurple}{$]$}}
 \end{tabular}
}

\newenvironment{singlebracketsr}[1][.4\linewidth]
{\begin{tabular}{m{.1cm}m{#1}m{.1cm}}
   \scalebox{1}[1]{\textcolor{DPurple}{$[$}} &
                                               \vspace{-.4cm}
                                               \hspace{-.4cm}
  \begin{minipage}{\linewidth}
    \begin{tabular}{p{\linewidth}}}
{\end{tabular}
\end{minipage}
   & \hspace{-.2cm}
     \scalebox{1}[1]{\textcolor{DPurple}{$\rangle$}}
 \end{tabular}
}

\newenvironment{ssinglebracketsr}[2][\linewidth]
{\def\relbracketscale{#2}
 \begin{tabular}{m{.1cm}m{#1}m{.1cm}}
   \scalebox{1}[\relbracketscale]{\textcolor{DPurple}{$[$}} &
                                               \vspace{-.4cm}
                                               \hspace{-.4cm}
  \begin{minipage}{\linewidth}
    \begin{tabular}{p{\linewidth}}}
{\end{tabular}
\end{minipage}
   & \hspace{-.2cm}
     \scalebox{1}[\relbracketscale]{\textcolor{DPurple}{$\rangle$}}
 \end{tabular}
}

\begin{document}

\title{KestRel: Relational Verification using E-Graphs for Program Alignment}

\author{Robert Dickerson}
\orcid{0000-0002-2697-2145}
\affiliation{%
  \institution{Purdue University}
  \city{West Lafayette}
  \country{USA}
}
\email{dicker18@purdue.edu}

\author{Prasita Mukherjee}
\orcid{0009-0001-4833-1496}
\affiliation{%
  \institution{Purdue University}
  \city{West Lafayette}
  \country{USA}
}
\email{mukher39@purdue.edu}

\author{Benjamin Delaware}
\orcid{0000-0002-1016-6261}
\affiliation{%
  \institution{Purdue University}
  \city{West Lafayette}
  \country{USA}
}
\email{bendy@purdue.edu}

\begin{abstract}
Many interesting program properties involve the execution of
\textit{multiple} programs, including observational equivalence,
noninterference, co-termination, monotonicity, and idempotency. One
strategy for verifying such \textit{relational properties} is to
construct and reason about an intermediate program whose correctness
implies that the individual programs exhibit those properties. A key
challenge in building an intermediate program is finding a good
\emph{alignment} of the original programs. An alignment puts subparts
of the original programs into correspondence so that their
similarities can be exploited in order to simplify verification. We
propose an approach to intermediate program construction that uses
e-graphs, equality saturation, and algebraic realignment rules to
efficiently represent and build programs amenable to automated
verification. A key ingredient of our solution is a novel data-driven
extraction technique that uses execution traces of candidate
intermediate programs to identify solutions that are semantically
well-aligned. We have implemented a relational verification engine
based on our proposed approach, called \kestrel{}, and use it to
evaluate our approach over a suite of benchmarks taken from the
relational verification literature.
\end{abstract}



\maketitle

\section{Introduction}
\label{sec:intro}

Program verification tools have matured considerably in recent years,
to the point that many mainstream programming languages are targeted
by at least one verification tool~\cite{Verus, gurfinkel2015,
  vst2018appel, frama-c}, and verification-aware languages now support
a rich set of features~\cite{Dafny, f-star, viper}. These tools and
languages are designed to automate verification of semantically rich
program properties, focusing on \emph{individual} program executions;
for example, showing that every final state of a program satisfies a
desired postcondition. Many interesting program behaviors, however,
involve the executions of \emph{multiple} programs. To verify
correctness of a program optimization, we must show that an original
program $p_1$ and its optimized version $p_2$ are
\textit{observationally equivalent}. That is, when $p_1$ and $p_2$ are
executed in the same initial state, they arrive at the same final
state. Proving this sort of \emph{relational} behavior requires
reasoning jointly about the executions of \emph{both} $p_1$ and
$p_2$. A variety of important program behaviors are inherently
relational~\cite{Hyperproperties, barthe2011}, including refinement,
idempotence, non-interference, and co-termination.

Somewhat surprisingly, reasoning about relational properties does not
require the development of specialized tooling, as most common
relational verification problems can be immediately reduced to
non-relational ones. This reduction is straightforward: given a pair
of target programs, we can construct an \textit{intermediate program}
that encodes their joint execution by renaming any shared variables
and concatenating the programs together~\cite{francez1983,
barthe2011secure}. The desired relational property can then be
established by applying single-program verification techniques to this
intermediate program. While this approach is theoretically sound, the
concatenated intermediate program is often prohibitively difficult to
verify in practice.

To demonstrate the limitations of this strategy, consider the pair of
programs, \lstinline|p1| and \lstinline|p2|, shown in
\autoref{fig:example}. Both programs iterate over a list of employees,
scheduling bonus payments for the identified workers via some
black-box financial services API. Program \lstinline|p2| does so
slightly more efficiently than \lstinline|p1|, however, as it caches
part of the bonus calculation prior to entering the loop. To establish
that this optimization is safe, we need to prove that, starting from
the same initial state, \lstinline|p1| and \lstinline|p2| schedule the
same set of payments. Since the two programs already operate over
disjoint variables, it suffices to show that
\lstinline[escapeinside={\$}{\$}]|payments$_1$| and
\lstinline[escapeinside={\$}{\$}]|payments$_2$| are the same after
executing the program \lstinline|p1; p2|.

\begin{figure}[!t]
\vspace{-.5cm}
\begin{center}
  \hfill
  \begin{subfigure}[t]{.5\textwidth}
\begin{lstlisting}[escapeinside={\$}{\$}]
int i$_1$ = 0;
while (i$_1$ < length(bonuses$_1$)) {
 int id$_1$ = bonuses$_1$.get(i$_1$);
 int sal$_1$ = emp$_1$.getSalary(id$_1$);
 payments$_1$.schedule(id$_1$, sal$_1$ * calc_bonus(rate$_1$));
 i$_1$ += 1; }
\end{lstlisting}
\end{subfigure}
\hfill
\rulesep
\hfill
\begin{subfigure}[t]{.47\textwidth}
  \begin{lstlisting}[escapeinside={\$}{\$}]
int i$_2$ = 0; int bonus$_2$ = calc_bonus(rate$_2$);
while (i$_2$ < length(bonuses$_2$)) {
 int id$_2$ = bonuses$_2$.get(i$_2$);
 int sal$_2$ = emp$_2$.getSalary(id$_2$);
 payments$_2$.schedule(id$_2$, sal$_2$ * bonus$_2$);
 i$_2$ += 1; }
\end{lstlisting}
\end{subfigure}
\hfill

\begin{subfigure}[b]{.47\textwidth}
  \centering
  \small
  \lstinline|p|$_1$
\end{subfigure}
\hfill
\begin{subfigure}[b]{.47\textwidth}
  \centering
  \small
  \lstinline|p|$_2$
\end{subfigure}
\end{center}
\vspace{-.2cm}
\caption{Two programs for calculating employee bonuses.}
\label{fig:example}
\vspace{-1.5em}
\end{figure}

In theory, we could do so using any verifier for the source language
of \lstinline|p1| and \lstinline|p2|. However, this task is beyond the
capabilities of many automated program verifiers, as reasoning about
this intermediate program requires a pair of loop invariants that
\emph{completely} characterize how each loop in \lstinline|p1; p2|
mutates its copy of \lstinline|payments|.  Even given such loop
invariants, establishing that they hold requires a detailed
specification of the \lstinline|schedule| method. In essence, we have
reduced the problem of showing this optimization is correct to proving
full functional correctness of \lstinline|p1| and
\lstinline|p2|~\cite{francez1983}. As this example illustrates, the
simple strategy of constructing an intermediate program by
concatenation can result in a disproportionately difficult
verification task. In fact, depending on the input programs, the loop
invariants and function specifications required to reason about the
intermediate program may not be expressible in a decidable
logic~\cite{shemer2019}.

\begin{wrapfigure}{r}{0.45\textwidth + 2\FrameSep +
    2\FrameRule\relax}
  \vspace{-1.5em}
\begin{lstlisting}[escapeinside={\$}{\$}]
int i$_1$ = 0; int i$_2$ = 0;
int bonus$_2$ = calc_bonus(rate$_2$);
while (i$_1$ < length(bonuses$_1$)) {
 int id$_1$ = bonuses$_1$.get(i$_1$);
 int id$_2$ = bonuses$_2$.get(i$_2$);
 int sal$_1$ = emp$_1$.getSalary(id$_1$);
 int sal$_2$ = emp$_2$.getSalary(id$_2$);
 payments$_1$.schedule(id$_1$, sal$_1$*calc_bonus(rate$_1$));
 payments$_2$.schedule(id$_2$, sal$_2$*bonus$_2$);
 i$_1$ += 1;  i$_2$ += 1; }
\end{lstlisting}
\begin{subfigure}[b]{.47\textwidth}
  \centering
  \small
  \lstinline|p3|
\end{subfigure}
\vspace{-.3cm}
\caption{A program that combines  \lstinline|p|$_1$ and
  \lstinline|p|$_2$ from  \autoref{fig:example}}
\label{fig:example-prod}
\vspace{-.3cm}
\end{wrapfigure} %

One solution to this problem is to find an intermediate program that
\textit{aligns} the original programs in a way that that better
captures the commonalities between their
subparts~\cite{barthe2011}. Better alignments enable verifiers to
exploit these similarities, simplifying the proof effort. To see how,
consider the program \lstinline|p3| in \autoref{fig:example-prod} to
the right, which also encodes the semantics of \lstinline|p1| and
\lstinline|p2|. In contrast to \lstinline|p1; p2|, \lstinline|p3|
features a single loop that encodes the simultaneous, or
\textit{lockstep}, execution of the loops in \lstinline|p1| and
\lstinline|p2|. This single loop captures the fact that each iteration
of these loops schedules the same payment, as they both call
\lstinline|payments.schedule($\ldots$)| with identical arguments. This
property is straightforwardly captured by a simple loop invariant
expressible in the theory of equality with uninterpreted functions
(EUF). In addition, rather than a full specification of
\lstinline|schedule|, we need only know that calling it with the same
arguments yields the same result, a property that can also be captured
in EUF. Importantly, EUF is supported by all modern SMT solvers,
unlocking the possibility of tractable automated verification. As this
example suggests, more sophisticated strategies for constructing
aligned intermediate programs can greatly simplify relational
reasoning.

This paper presents an automated approach to intermediate program
construction which aligns a pair of target programs in order to
effectively reason over their joint execution. While strategies exist
for constructing the kind of intermediate program given in
\lstinline|p3|, they tend to operate
\emph{syntactically}~\cite{ZP+08} by identifying locations or
\textit{cut points} in the input programs to bring into
alignment. This places strong constraints on the shape of the input
programs, effectively requiring a one-to-one correspondence between
program locations. Such approaches fail to take into account how the
\textit{semantics} of programs affect their alignment.
\begin{wrapfigure}{r}{\dimexpr 0.37\textwidth + 2\FrameSep +
    2\FrameRule\relax}
  \vspace{-.3cm}
  \begin{subfigure}[t]{.18\textwidth}
    \centering
\begin{lstlisting}
int z1 = 0;
int y1 = 0;
z1 = 2*x1;
while (z1>0) {
  z1 = z1 - 1;
  y1 = y1 + x1;
}
\end{lstlisting}
    \lstinline|double1|
    \vspace{-.2cm}
  \end{subfigure}
  \hfill
  \rulesep
  \hfill
  \begin{subfigure}[t]{.18\textwidth}
    \centering
\begin{lstlisting}
int z2 = 0;
int y2 = 0;
z2 = x2;
while (z2>0) {
  z2 = z2 - 1;
  y2 = y2 + x2; }
y2 *= 2;
\end{lstlisting}
    \lstinline|double2|
    \vspace{-.2cm}
  \end{subfigure}
  \caption{A good alignment matches every iteration of the loop in
    \lstinline|double1| with two iterations of the loop in
    \lstinline|double2|.}
  \label{fig:stutter+loop}
\vspace{-.3cm}
\end{wrapfigure}
Consider the pair of programs in \autoref{fig:stutter+loop} taken from
\citet{unno2021}. Each program sets its version of \lstinline!y! to
\lstinline!2x!, but the loop in \lstinline|double1| executes twice as
many times as the loop in \lstinline|double2|. A single loop that
encodes the behaviors of both loops must align each iteration of the
loop in \lstinline|double2| with two iterations of the one in
\lstinline|double1|. One challenge when considering these kinds of
semantic alignments is that they naturally lead to larger sets of
candidate intermediate programs. While the number of ways to align the
locations in a pair of programs is already large, $O(|p_1| \cdot
|p_2|)$, the space of semantic alignments involving these kinds of
loop schedulings and unrollings is potentially unbounded; in this
example, an arbitrary number of iterations of the loop in
\lstinline|double1| can be aligned with an arbitrary number of
iterations of the loop in \lstinline|double2|. Even if we limit
ourselves to matching a bounded number $n$ of iterations between
loops, there are now $O(n \cdot n)$ ways of aligning \emph{every} pair
of loops in the target programs. A successful semantic alignment
strategy must be able to efficiently represent and explore a large
(possibly infinite) set of candidate intermediate programs.

Our solution to this challenge is to use e-graphs~\cite{nelson1980,
  nieuwenhuis2005, willsey2021} to compactly represent a space of
semantically aligned intermediate programs. While e-graphs have
previously been used to efficiently represent and explore sets of
equivalent programs in order to find one with the best
performance~\cite{tate2009}, we use them to identify intermediate
programs that are amenable to verification. Instead of representing
these programs directly, we instead embed them in a relational
calculus equipped with algebraic \emph{realignment rules} in the
spirit of \citet{antonopoulos2023}. This allows us to frame the search
for a good alignment as a search for a set of realignment rule
applications starting from a na\"{i}ve concatenative embedding of the
original program. To compactly represent this space of rule
applications, we initialize an e-graph with the na\"{i}ve alignment
and saturate it with the set of realignment rules. A key invariant of
the resulting e-graph is it only contains elements corresponding to
intermediate programs that are semantically equivalent to the original
pair of programs (\autoref{thm:reify+preserves+eqv}). A pleasant
consequence of our strategy is that it can easily incorporate existing
semantics-preserving transformations on individual program, e.g., loop
unrolling, that can unlock better alignments.

As the name suggests, finding a good semantic alignment depends on the
\emph{semantics} of the target programs. Thus, our approach identifies
promising alignments by examining concrete program traces, generated
by executing candidate intermediate programs. As an example, loops in
the intermediate program that combine loops from the input programs,
e.g., \lstinline|p3|, are likely not well-aligned unless the loop
conditions from each input program become false at the same time. When
exploring the space of alignments, we can favor candidate programs
whose execution traces exhibit this property. Importantly, our
approach always produces a semantically equivalent intermediate
program, even when no finite set of traces is capable of capturing the
relationship between the input programs. This is in contrast to the
data-driven approach of \citet{churchill2019}, which is only
guaranteed to find programs that cover a set of test cases.
\begin{wrapfigure}{r}{\dimexpr 0.45\textwidth + 2\FrameSep +
    2\FrameRule\relax}
  \vspace{-.3cm}
  \begin{subfigure}[t]{.22\textwidth}
\begin{lstlisting}
int f(int m, int n) {
  int k := 0;
  for i in 0..n {
    k += m;
  }
  return k;
}
\end{lstlisting}
    \vspace{-.4cm}
  \end{subfigure}
  \hfill
  \rulesep
  \hfill
  \begin{subfigure}[t]{.23\textwidth}
\begin{lstlisting}
int g(int m, int n) {
  int k := 0;
  for i in 0..n {
    for j in 0..m {
      k++;
  }}
  return k; }
\end{lstlisting}
    \vspace{-.3cm}
  \end{subfigure}
  \caption{The function \lstinline|f| optimizes \lstinline|g| by
    collapsing its inner loop into a single addition operation.}
\label{fig:data+dependent+loop}
  \vspace{-1em}
\end{wrapfigure}
Our approach is able to find an intermediate program with a desirable
alignment for the programs in \autoref{fig:data+dependent+loop}, which
\citet{churchill2019} identifies as particularly challenging. To the
best of our knowledge, no other existing technique is capable of
building an aligned intermediate program for these programs.

To demonstrate the feasibility of our approach, we have built a
prototype relational verification tool, \kestrel{}. \kestrel{} targets
a subset of C and supports Dafny~\cite{Dafny} and
SeaHorn~\cite{gurfinkel2015} as backends, enabling it to verify
relational properties of programs that use APIs to manage abstract
data types with hidden internal state, as well as array-manipulating C
programs.  We have used \kestrel{} to produce aligned intermediate
programs for a diverse suite of benchmarks and relational properties
taken from the literature, including examples that fall outside the
reach outside of similar tools. Our experimental results show that
\kestrel{} discovers alignments that enable verification to succeed
where simpler alignment strategies fail.

We pause here to note that an alternative strategy is to instead
develop bespoke relational verification tools tailored to assertions
over multiple program executions~\cite{francez1983, shemer2019,
  unno2021, itzhaky2024, farzan2019automated,
  farzan2019reductions}. These tools also try to exploit syntactic and
semantic commonalities between the target programs in order to
simplify the verification task.\footnote{What we refer to as
  ``intermediate programs'' in this work are often called ``product
  programs'' in the literature. However, many bespoke relational
  verification approaches use the term ``product program'' to refer to
  their embeddings of multiple programs. This is fundamentally
  distinct from our notion of intermediate programs as standalone
  programs that are written in the same language as the input programs
  and that are independent of a specific verification technique. To
  avoid confusion, we use ``intermediate program'' for the latter
  connotation of ``product program'' throughout this paper.}  A
popular example of this approach is \textit{relational program
  logics}, which adapt the judgment of traditional program
logics~\cite{hoare1969}, to range over multiple programs
~\cite{francez1983, benton2004, Yang+07, BKOZ+13,
  banerjee2016, aguirre2017, MHRV+19}. These logics are equipped with
specialized rules for reasoning about the joint behaviors of programs.
Many logics include a specialized rule for reasoning about
``lockstep'' loops whose iterations are in a perfect 1-1
correspondence, for example, enabling much simpler loop
invariants~\cite{sousa2016}. While other bespoke relational
verification tools differ in their alignment strategies, they all
attempt to exploit commonalities between the target programs to
simplify the verification task. 
\autoref{sec:related} provides a more detailed discussion of these
approaches.

In summary, this paper describes the following contributions:
\begin{itemize}
\item We present a novel application of e-graphs to build and
  compactly represent the space of aligned intermediate programs
  expressed in a domain of relational alignments equipped with
  algebraic realignment rules.
\item We develop a hybrid extraction technique that combines a
  syntactic cost metric with a novel non-local extraction technique
  that uses dynamic execution traces to identify alignments amenable
  to automated verification.
\item We present a relational verification framework, \kestrel{}, that
  implements this approach, and demonstrate its utility by evaluating
  it on a diverse set of challenging relational verification
  benchmarks drawn from the literature.
\end{itemize}

The remainder of the paper is structured as follows. We begin with a
brief primer on e-graphs, followed by an overview of our
approach. \autoref{sec:formalization} then formalizes our approach to
relational verification using a core calculus for relational
alignment. Next, \autoref{sec:alignment} describes how we use e-graphs
to construct and represent a set of candidate intermediate programs,
and then presents our data-driven technique for identifying and
extracting an intermediate program amenable to automated verification
from this space.  \autoref{sec:evaluation} presents an empirical
evaluation of \kestrel{}, a relational verification tool based on our
approach. The paper ends with related work and conclusions.

\section{Overview}
\label{sec:overview}

\begin{figure}
  \includegraphics[width=0.65\textwidth,angle=0]{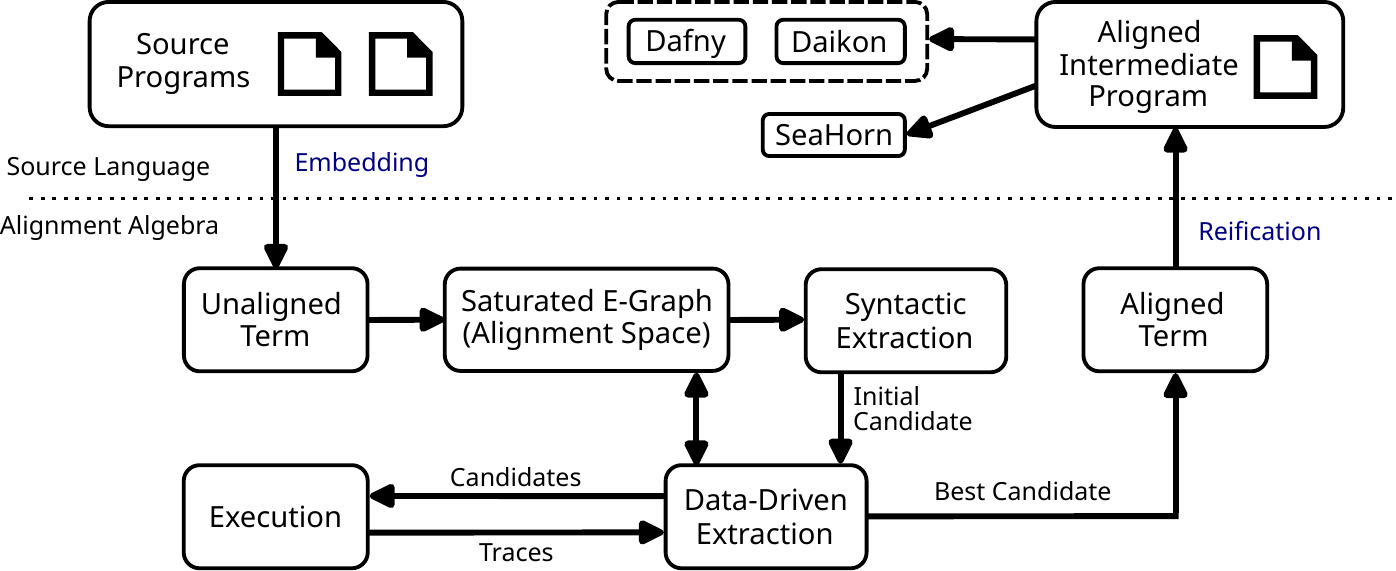}
  \caption{High-level overview of \kestrel{}.}
  \label{fig:overview}
  \vspace{-1em}
\end{figure}

\autoref{fig:overview} presents a high-level overview of our approach
to relational verification. Given a pair of programs as input, our
goal is to output a single intermediate program which can be handed
off to automated, non-relational verification tools. To do so, we
first embed the input programs as a term in a relational algebra and
insert this term into an e-graph. We construct a space of aligned
programs by saturating this e-graph using a set of algebraic
realignment rules. Using purely syntactic criteria (e.g., the number
of fused loops in the alignment), we identify an initial candidate. We
then give this initial alignment to a data-driven extraction method
which uses the e-graph to look for better alignments, examining
execution traces to measure the quality of a candidate alignment. The
best alignment from this process is then reified into a single
intermediate program. This program can then be handed to an
off-the-shelf verifier like Dafny or SeaHorn.

\subsection{Introduction to E-Graphs}

We begin with a brief overview of e-graphs and equality saturation;
readers familiar with these topics may safely skip this section.

\paragraph{E-Graphs}
An e-graph~\cite{nieuwenhuis2005, nelson1980} is a data structure
which compactly represents equivalence classes on sets of terms. An
e-graph contains \textit{e-nodes} and \textit{e-classes}, where each
e-node associates a symbol with a (possibly empty) list of child
e-classes and each e-class contains a collection of equivalent
e-nodes. Each e-class has a unique identifier, and an equivalence
relation between e-class identifiers is stored in an 
union-find structure~\cite{tarjan1975}.

\begin{definition}[E-Graph]
An \textit{e-graph} is a triple $(U, M, H)$ where:
\begin{itemize}
  \item $U$ is a union-find data structure over e-class identifiers.
  \item $M$ is a mapping from e-class identifiers to to e-classes such
that all identifiers equivalent in $U$ map to the same e-class:
$\forall i,j.\ M[i] = M[j] \iff \textsf{find}(U, i) = \textsf{find}(U,
j)$.
  \item $H$ is a mapping from e-nodes to e-class identifiers.
\end{itemize}
\end{definition}

A term $t$ is \textit{represented} in an e-graph or e-class if there
exists an e-node $d$ in the e-graph or e-class such that:
\begin{enumerate}
\item $d$'s symbol matches the symbol at $t_r$, the root of $t$'s
  syntax tree,
\item $d$ has the same number of children as $t_r$, and
\item each child $t_0, \ldots, t_n$ of $t_r$ is represented by a
  corresponding child e-class $c_0, \ldots, c_n$ of $d$.
\end{enumerate}

\begin{figure}[t]
  \begin{subfigure}[t]{.26\textwidth}
    \centering
  \includegraphics[width=0.7\textwidth,angle=0]{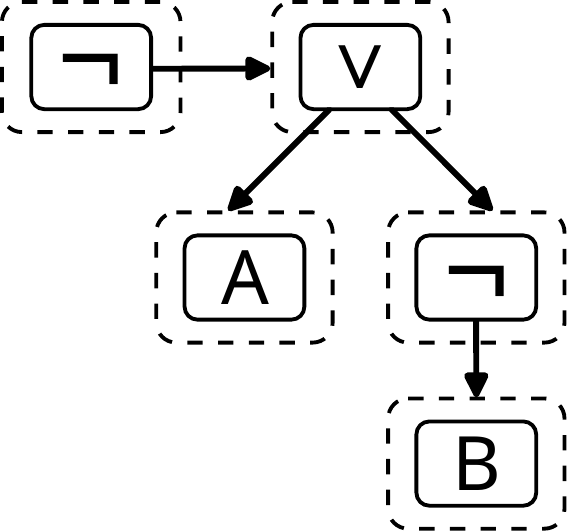}
  \caption{An e-graph containing only $\neg (A \lor \neg B)$.}
  \label{fig:egraph-examples-a}
\end{subfigure}
\hfill
\begin{subfigure}[t]{.33\textwidth}
  \centering
  \includegraphics[width=0.7\textwidth,angle=0]{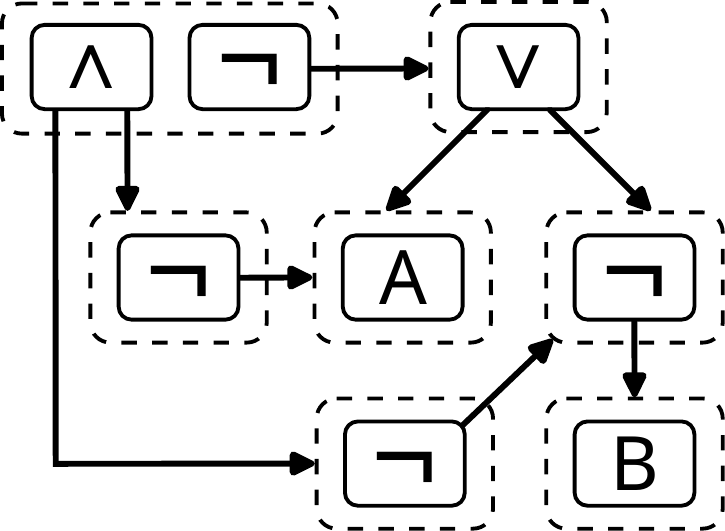}
  \caption{E-graph (a) with $\neg(x \lor y) \rightarrow$ \\ $\neg x \land
\neg y$ applied.}
  \label{fig:egraph-examples-b}
\end{subfigure}
\hfill
\begin{subfigure}[t]{.33\textwidth}
  \centering
  \includegraphics[width=0.7\textwidth,angle=0]{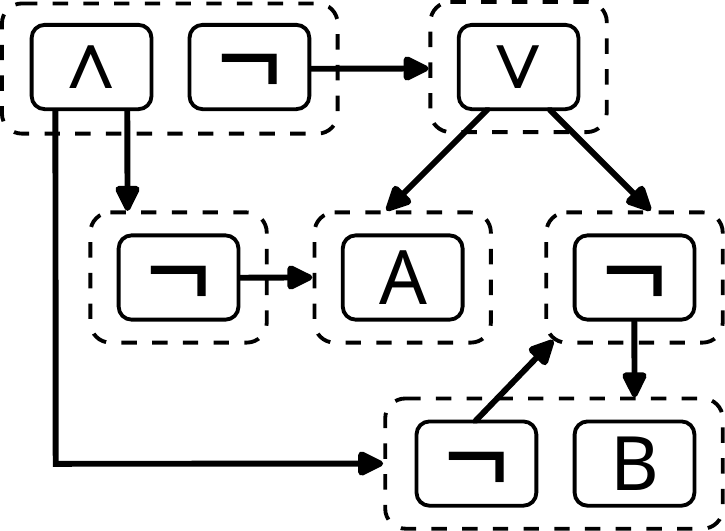}
  \caption{E-graph (b) with $\neg\neg x$ $\rightarrow x$ applied.}
  \label{fig:egraph-examples-c}
\end{subfigure}
\vspace{-.5em}
\caption{Example e-graphs.}
\vspace{-.5em}
\end{figure}

\begin{example}
  Using classical Boolean logic as an example language,
  \autoref{fig:egraph-examples-a} shows the term
  $\neg (A \lor \neg B)$ represented as an e-graph. E-nodes and
  e-classes are displayed as solid and dashed boxes, respectively. Our
  initial e-graph has e-nodes for each node in the term's syntax tree,
  and each of these e-nodes is in a singleton e-class. While each
  e-class in this e-graph contains a single e-node, note that the
  children of e-nodes are \textit{e-classes}, not other e-nodes.  To
  extract a term represented by an e-graph, we start at its root
  e-class and recursively chose an e-node from its child e-classes. In
  this case, each e-class has exactly one choice, so this e-graph only
  represents the original term.
\end{example}

\paragraph{Rewrites Rules}
E-graphs can compactly represent the term equivalences induced by a
set of rewrite rules.  Rewrite rules have the form $l \rightarrow r$,
which intuitively says any subterm of an e-graph that matches $l$ may
be replaced by the term $r$. Both $l$ and $r$ can contain variables;
these are instantiated when the rule is applied. An example of a sound
rewrite rule in Boolean algebra is $x \lor \bot \rightarrow x$, which
says a disjunction of false with any term $x$ is equivalent to $x$.

To add representations of terms rewritten according to some rule $l
\rightarrow r$ to an e-graph, we:
\begin{enumerate}
\item locate all pairs $(c_i, \sigma_i)$ where $c_i$ is an e-class
  representing a term that matches $l$ and $\sigma_i$ is an
  appropriate instantiation of the variables in $l$,
\item for each $c_i$, add a new e-class $e_i$ to the e-graph such that
  $e_i$ represents $r[\sigma_i]$, where $r[\sigma_i]$ is $r$ with
  substitutions in $\sigma_i$ applied, and finally
\item merge each $c_i$ with each $e_i$.
\end{enumerate}
The first task is accomplished using
\textit{e-matching}~\cite{e-matching}, while the second and third are
handled by the \texttt{add} and \texttt{merge} operations of the
e-graph's union-find structure $U$.

\begin{example}
  One sound rewrite rule we can apply in the context of classical
  Boolean logic is given by De Morgan's laws:
  $\neg (x \lor y) \rightarrow \neg x \land \neg y$. The upper left
  e-class in \autoref{fig:egraph-examples-a} has a match with the
  left-hand side of this rewrite rule where
  $\sigma = [x \mapsto A, y \mapsto \neg
  B]$. \autoref{fig:egraph-examples-b} shows the e-graph in
  \autoref{fig:egraph-examples-a} with this rewrite applied; a new
  $\land$ e-node has been created and merged with the original
  matching e-class. We can extract the rewritten term
  $\neg A \land \neg \neg B$ from this e-graph by choosing the $\land$
  e-node as the representative of the root e-class, and recursively
  selecting the singleton e-nodes as the representatives of each of
  its descendants.
\end{example}

\begin{example}
  Another sound rewrite rule in classical Boolean logic is double
  negation elimination: $\neg \neg x \rightarrow x$.
  \autoref{fig:egraph-examples-c} depicts
  \autoref{fig:egraph-examples-b} with this rewrite applied. The
  e-nodes corresponding to the terms $\neg \neg B$ and $B$ are now
  equivalent as the two bottommost e-classes have been merged. We can
  find the term $\neg A \land B$ by starting with $\land$ in the upper
  left e-class as before, choosing $B$ for its first child in the
  lower right e-class, and choosing the only available e-node in all
  other e-classes. Note also that the e-graph in
  \autoref{fig:egraph-examples-c} represents an unbounded number of
  terms of the form ${(\neg \neg)}^\ast B$, where $^\ast$ is Kleene
  star, as we can traverse the cycle between the bottom right e-classes
  an arbitrary number of times.
\end{example}

\paragraph{Equality Saturation and Extraction}
Each application of a rewrite rule to an e-graph is additive,
potentially creating new e-nodes and merging e-classes, but continuing
to represent all previously represented terms. \textit{Equality
saturation}~\cite{tate2009} is the process of initializing an e-graph with
some term, and then continually matching and applying rewrite rules to
that e-graph until either no new opportunities for rewrites are found
or some bound or timeout is reached. The result is an e-graph which
represents a (potentially infinite) space of terms which are
equivalent to the original under some sequence of rewrite rule
applications.

Once an e-graph has been saturated in this way, the challenge is how
to select or \textit{extract} the best term, according to some metric.
Metrics that are defined in terms of subterms often admit a simple
extraction technique. In order to extract the smallest representative
term in an e-graph, for example, we first recursively assign each
e-node a \textit{cost} of one more than the sum of the costs of its
child e-class, and each e-class the minimum cost of its constituent
e-nodes. Extraction then proceeds in a top-down manner by selecting
the lowest cost e-node in each e-class. More complicated metrics may
require more sophisticated extraction techniques. In the case of
\kestrel{}, the quality of a program alignment depends on the
semantics of the intermediate program it represents; this is the
motivation for our data-driven approach to extraction.

\subsection{Example \kestrel{} Workflow}

\begin{wrapfigure}{r}{\dimexpr 0.35\textwidth + 2\FrameSep +
    2\FrameRule\relax}
  \vspace{-1cm}
  \centering
\begin{lstlisting}
int y1 = 0; int y2 = 0;
int z1 = 2*x1; int z2 = x2;
while (z2 > 0)
  { z1--; y1 += x1; z1--; y1 += x1;
    z2--; y2 += x2 }
y2 *= 2;
\end{lstlisting}
  \lstinline|double3|
  \vspace{-.5em}
\caption{An intermediate program encoding the behaviors of
  \lstinline|double1| and \lstinline|double2|. }
\label{fig:intermediate+double}
\vspace{-1em}
\end{wrapfigure}
We illustrate the key pieces of our proposed approach to intermediate
program construction by showing how we build \lstinline|double3| shown
in \autoref{fig:intermediate+double}, given \lstinline|double1| and
\lstinline|double2| from \autoref{fig:stutter+loop}. Importantly,
verifying that \lstinline|y1 = y2| after executing \lstinline|double3|
requires a loop invariant that is a simple equality between the values
of $\lstinline|y1| = 2*\lstinline|y2|$, while verifying
\lstinline!double1; double2! requires two loop invariants, each of
which involve \lstinline|x|, \lstinline|y|, and
\lstinline|z|.

\begin{figure}
\begin{minipage}{.39\linewidth}
  \begin{relbrackets}[8]{0.5}{0.5}
  \lstinline|int y1 = 0;|        & \lstinline|int y2 = 0;| \\
  \lstinline|int z1 = 2 * x1;|   & \lstinline|int z2 = x2;| \\
  \lstinline|while (z1 > 0) {|   & \lstinline|while (z2 > 0) {| \\
  \lstinline|  z1--; y1 += x1 }| & \lstinline|  z1--; y1 += x1 }| \\
                                 & \lstinline|y2 *= 2| \\
  \end{relbrackets}
\end{minipage}
\hspace{1cm}
$\lstinline|==|$
\begin{minipage}{.16\linewidth}
  \begin{ssinglebracketsl}{8}
  \lstinline|int y1 = 0;| \\
  \lstinline|int z1 = 2 * x1;| \\
  \lstinline|while (z1 > 0) {| \\
  \lstinline|  z1--; y1 += x1 }| \\
  \end{ssinglebracketsl}
\end{minipage}
\hspace{0.8cm}
\lstinline|;;|
\hspace{-0.4cm}
\begin{minipage}{.16\linewidth}
\begin{ssinglebracketsr}{8}
  \lstinline|int y2 = 0;| \\
  \lstinline|int z2 = x2;| \\
  \lstinline|while (z2 > 0) {| \\
  \lstinline|  z1--; y1 += x1 }| \\
  \lstinline|y2 *= 2| \\
\end{ssinglebracketsr}
\end{minipage}
\hspace{1cm}
$\lstinline|==|$
\vspace{0.5cm}
\hspace{1.5cm}
\hfill
\\
\hspace{-1cm}
$\cdots\lstinline|==|$
\hspace{-0.2cm}
\begin{minipage}{.35\linewidth}
  \begin{tabular}{l}
    \begin{relbrackets}[2.5]{0.5}{0.5}
    \lstinline|int y1 = 0|        & \lstinline|int y2 = 0;| \\
    \lstinline|int z1 = 2 * x1;|  & \lstinline|int z2 = x2| \\
    \end{relbrackets} \hspace{-0.4cm}\lstinline|;;|\vspace{0.2cm}
    \\
    \begin{relbrackets}[4]{0.5}{0.5}
    \lstinline|while (z1 > 0) {|  & \lstinline|while (z2 > 0) { |\\
    \lstinline|  z1--;|           & \lstinline|  z2--; |\\
    \lstinline|  y1 = y1 + x1 }|  & \lstinline|  y2 = y2 + x2 }| \\
    \end{relbrackets} \hspace{-0.45cm} \lstinline|;;|
    \\
    \hspace{0.2cm}
    \begin{singlebracketsr}[0.8cm]
    \lstinline|y2 *= 2|
    \end{singlebracketsr}
  \end{tabular}
\end{minipage}
\hspace{1.9cm}
$\lstinline|==|$
\hspace{-0.4cm}
\begin{minipage}{.35\linewidth}
  \begin{tabular}{l}
    \begin{relbrackets}[2.5]{0.55}{0.3}
    \lstinline|int y1 = 0|        & \lstinline|int y2 = 0;| \\
    \lstinline|int z1 = 2 * x1;|  & \lstinline|int z2 = x2| \\
    \end{relbrackets} \hspace{-0.4cm}\lstinline|;;|
    \vspace{-0.2cm}
    \\
    \hspace{0.4cm}
    \lstinline|whileSt 2 1|
    \begin{minipage}{.35\linewidth}
      \vspace{0.27cm}
      \begin{relbrackets}{0.5}{0.5}
      \lstinline|z1 > 0| & \lstinline|z2 > 0|
      \end{relbrackets}
    \end{minipage}
    \vspace{-0.2cm}
    \\
    \hspace{1.7cm}
    \begin{relbrackets}[2.5]{0.5}{0.01}
    \lstinline|z1--;|    & \lstinline|z2--;|\\
    \lstinline|y1 += x1| & \lstinline|y2 += x2|\\
    \end{relbrackets} \hspace{-0.45cm} \lstinline|;;|
    \vspace{0.1cm}
    \\
    \hspace{0.2cm}
    \begin{singlebracketsr}[0.8cm]
    \lstinline|y2 *= 2|
    \end{singlebracketsr}
  \end{tabular}
\end{minipage}
\hfill
\caption{Abbreviated derivation of an alignment using the rewrite
  rules presented in \autoref{sec:alignment}. The initial term relates
  two programs, both of which set $y$ to $2x$. The program on the left
  does this by counting to $2x$, while the program on the right counts
  to $x$ before multiplying by 2. The final term aligns the pre-loop
  initializations, the loop executions (with two iterations of the
  left program's loop for every one of the right's), and does not
  align the right-only \lstinline|y *= 2| with anything.}
\label{fig:derivation}
\vspace{-1em}
\end{figure}

\paragraph{An Algebra of Alignments}
Our first step in constructing \lstinline!double3! is to embed
\lstinline!double1! and \lstinline!double2! into a richer domain that
provides a more structured representation of intermediate programs. We
refer to elements of this domain as \emph{alignments} of (a pair of)
programs. The simplest alignment has the form \lstinline!<<p1|p2>>!;
this alignment represents an intermediate program which fully executes
\lstinline!p1! and then \lstinline!p2!, i.e. \lstinline!p1; p2!. Our
domain also includes finer-grained alignments that group together
subterms of the intermediate program. The alignment
\lstinline!<<s1 | t1; t2>>;; <<s2 | t3>>!, for example, groups
together the first statement of \lstinline|s1; s2| with the first two
statements of \lstinline|t1; t2; t3| and aligns the last statements of
both programs; these subalignments are composed together with the
$\relcomp{}$ operator. This domain is equipped with other relational
operators for aligning different control flow operators. The most
important of these is the %
\lstinline!whileR <<b1|b2>> <<c1|c2>>! operator, which represents an
intermediate program that executes the bodies of two loops in
lockstep. The final alignment in \autoref{fig:derivation} encodes
\lstinline|double3| using a variant of this operator,
\lstinline!whileSt m n <<b1|b2>> <<c1|c2>>!, which executes
\lstinline|c1 m| times and \lstinline|c2 n| times on each iteration.

Alignments are equipped with an equivalence relation, \lstinline|==|.
Intuitively, equivalent alignments represent semantically equivalent
intermediate
programs. This equivalence admits several relational \emph{realignment
  laws} which can be used to reason about the equivalence of different
alignments. The equivalence of all of the alignments shown in
\autoref{fig:derivation} are justified by these laws, for
example. Importantly, the alignment that encodes \lstinline|double3|
can be automatically derived from \lstinline!<<double1 | double2>>!
via a sequence of rewriting steps.

\begin{figure}[b]
\vspace{-1em}
\begin{tabular}{@{}c@{}c@{}}
  \includegraphics[width=0.4\textwidth,angle=0]{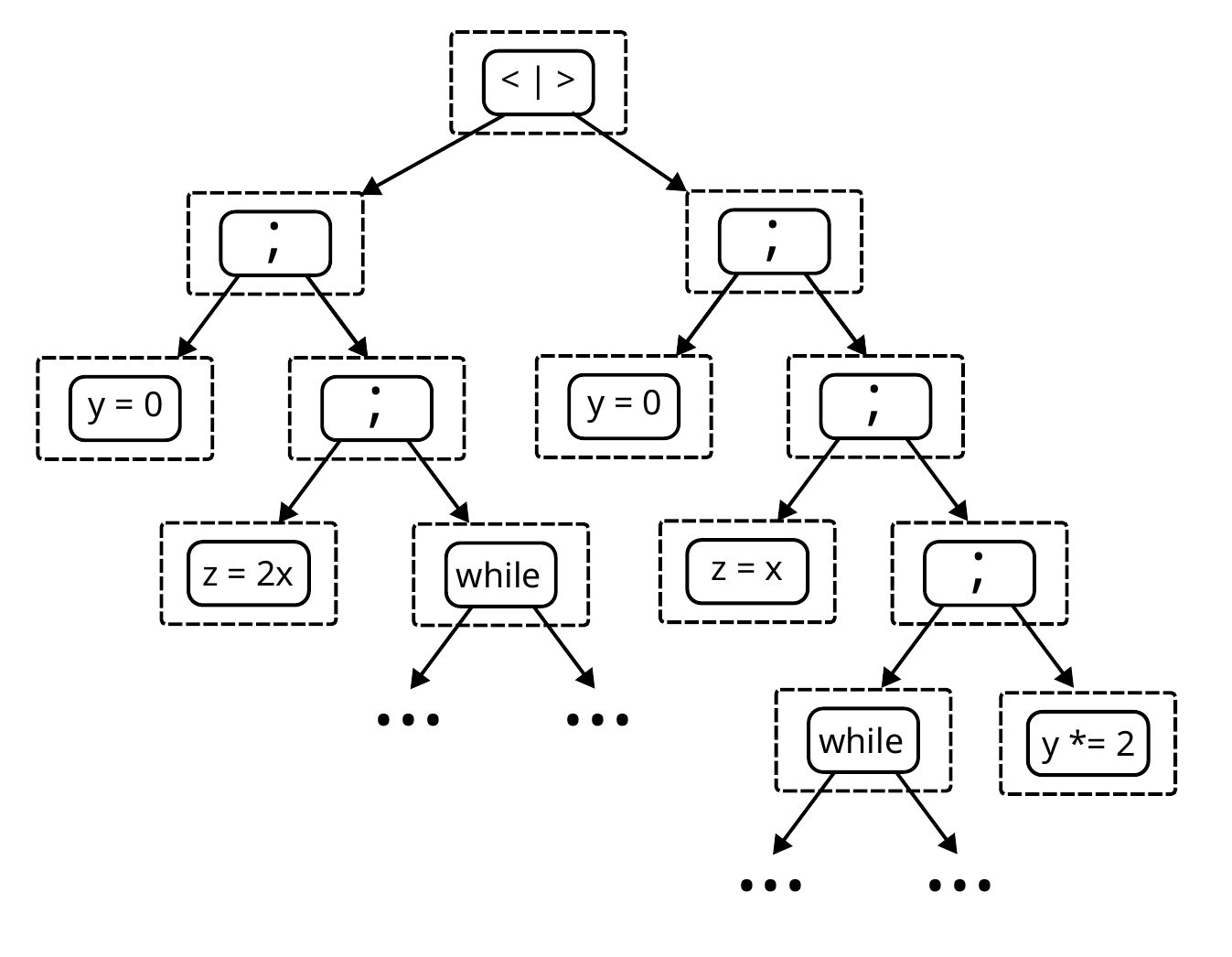}
  & \includegraphics[width=0.4\textwidth,angle=0]{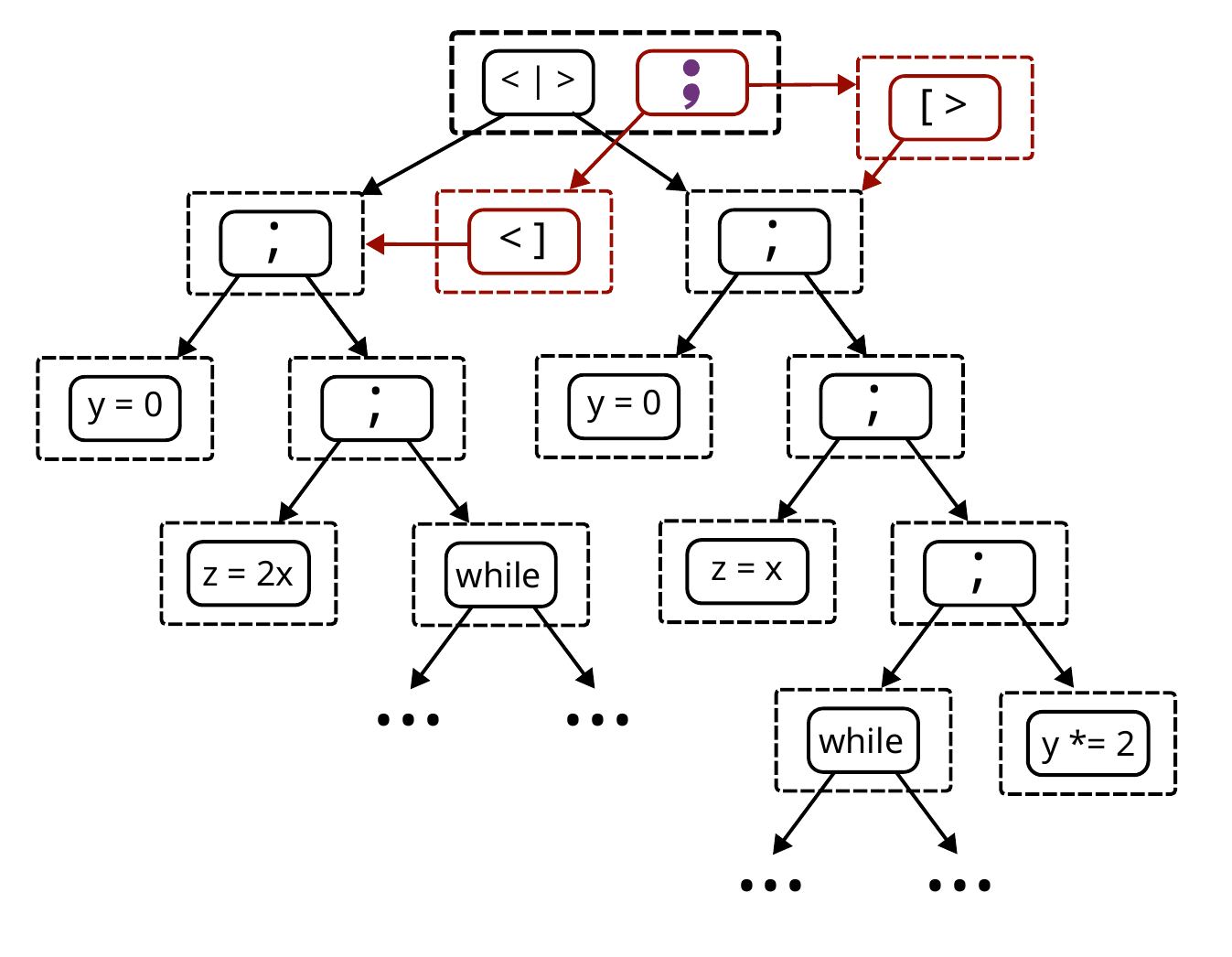} \\
  \vspace{-2em} \\
  (a) & (b)
\end{tabular}
\vspace{-1em}
 \caption{E-graphs containing representations of possible alignments
between \lstinline{double}$_1$ and \lstinline{double}$_2$. The e-graph
on the left (a) contains only the initial embedding. The e-graph on
the right (b) contains the initial embedding plus an application of
the \textsc{rel-def} given in \autoref{fig:rewritingRules}. (Added
nodes are depicted in red.) It is possible to extract both the first
and second terms in \autoref{fig:derivation} from (b).}
\label{fig:double-egraphs}
\vspace{-1em}
\end{figure}

\paragraph{Representing Possible Alignments with E-Graphs}
While the chain of rewrites shown in \autoref{fig:derivation} yields a
desirable alignment, many other equivalent alignments can be similarly
derived via realignment laws. To explore the set of equivalent
alignments, we use e-graphs as a compact representation of the space
of alignments. \autoref{fig:double-egraphs}(a) gives a simplified
representation of the \lstinline!<<double$_1$ | double$_2$>>! as an
e-graph, while \autoref{fig:double-egraphs}(b) depicts an e-graph that
simultaneously encodes both the first and second alignments in
\ref{fig:derivation}, as reflected by the inclusion of the
\lstinline!<<|>>! and \lstinline|;;| e-nodes in the same e-class.
Saturating the e-graph in \autoref{fig:double-egraphs}(a) with a set
of realignment rules results in an e-graph that includes the alignment
corresponding to \lstinline!double3!.

\paragraph{Searching for Desirable Alignments}
Once a fully saturated e-graph that represents the space of possible
alignments of \lstinline!<<double1 | double2>>! is in hand, our next
step is to \emph{extract} \lstinline!double3! from the set of
intermediate
programs embedded in the e-graph. Modern e-graph
libraries~\cite{willsey2021} are equipped with a mechanism that
greedily extracts terms by recursively using a cost function to select
the ``best'' representative of each equivalence class. This strategy
is inherently \textit{syntactic}, selecting nodes based on the terms
they represent. However, identifying the best alignment often involves
\textit{semantic} properties of the intermediate program it
represents. Finding the alignment that produces \lstinline!double3!,
for example, requires the observation that the body of the loop in
\lstinline|double1| must be executed twice for every execution of
\lstinline|double2|. (See \autoref{sec:comparingAlignments} for more
discussion on identifying desirable alignments.) To find alignments
with this kind of semantic property, we use a data-driven extraction
technique that examines traces of states generated by from candidate
intermediate program executions to determine alignment quality.
Observing dynamic traces allows our extraction mechanism to observe
this semantic relationship. We use a Markov-Chain Monte-Carlo
(MCMC)-based~\cite{hastings1970monte} algorithm to sample programs
from promising parts of the search space, using the e-graph to provide
neighboring extraction candidates during the search. Once a promising
candidate alignment has been found, our final step is to reify it into
an intermediate program, e.g., \lstinline|double3|, which can then be
given to an off-the-shelf single program verifier like
Dafny~\cite{Dafny} or SeaHorn~\cite{gurfinkel2015}.

\section{The \coreRel{} Calculus}
\label{sec:formalization}
\begin{figure}[t!]
  \begin{subfigure}[T]{0.4\textwidth}
    \vskip 0pt
  \begin{tabular}{R@{\hspace{2pt}}L@{\hspace{6pt}}}
    \footnotesize
    \lstinline|a|\production
    & \hfill \textsc{Integer Expressions} \\
    & \lstinline|n|
      \mid \lstinline|x|
      \mid \lstinline|a + a| \mid \lstinline|a - a|
      \mid \lstinline|a * a| \\
    \lstinline|b|\production
    & \hfill \textsc{Boolean Expressions} \\
    & \lstinline|true|
      \mid \lstinline|false|
      \mid \lstinline|a = a|
      \mid \lstinline|a < a| \\
    & \mid \lstinline|not b|
      \mid \lstinline|b /\ b| \\
    \lstinline|c|\production
    & \hfill \textsc{Commands} \\
    &  \lstinline|skip|
      \mid \lstinline|c; c|
      \mid \lstinline|x := a|
      \mid \lstinline|while b c| \\
    & \mid \lstinline|if b then c else c| \\[1mm]
  \end{tabular}
  \begin{tabular}{R@{\hspace{2pt}}L@{\hspace{6pt}}}
    \footnotesize
    \lstinline!if b then c :== if b then c else skip!
  \end{tabular}
\end{subfigure}
\quad \rulesep \quad
\begin{subfigure}[T]{0.5\textwidth}
  \vskip 0pt
  \footnotesize
  \begin{tabular}{R@{\hspace{2pt}}L@{\hspace{6pt}}}
    \lstinline|r|\production
    & \hfill \textsc{Aligned Commands}\\
    & \phantom{\mid} \lstinline!<< c | c >>! \\
    & \mid \lstinline!r ;; r!\\
    & \mid \lstinline!ifR << b | b >> then r else r !\\
    & \mid \lstinline!whileR << b | b >> r! \\ \\
  \end{tabular}

  \begin{tabular}{L@{\hspace{2pt}}L@{\hspace{6pt}}}
    \lstinline!<<s|] :== <<s | skip>>! \\
    \lstinline![|s>> :== <<skip | s>>! \\
    \lstinline!whileSt n m <<b1 | b2>> <<c1 | c2>> :==!\\
    \quad \lstinline!whileR <<b1 | b2>> <<!
    \overline{\text{\lstinline!if b1 then c1!}}^\texttt{n}
    \, \lstinline!|!\, \overline{\lstinline!if b2 then c2!}^\texttt{m}
    \lstinline!>>!
  \end{tabular}
\end{subfigure}
\caption{Syntax and notations for \imp{} and \coreRel{}.}
\label{fig:Syntax}
\vspace{-1em}
\end{figure}

This section describes a core calculus for program alignment, called
\coreRel{}, which we use to formalize our approach to intermediate
program construction.\footnote{The anonymized supplementary material
  includes a complete Coq formalization of \coreRel{} and its
  metatheory.} Our target programming language is the simple
imperative programming language, \imp{}, whose syntax is shown on the
lefthand side of \autoref{fig:Syntax}. The calculus assumes an
infinite set of identifiers for program variables; program states are
partial functions from identifiers to integers.  The semantics of an
\imp{} program \lstinline|c| is given by a completely standard
big-step reduction relation from input states $\sigma$ to output
states $\sigma'$: $\sigma, \lstinline|c| \Downarrow \sigma'$. \imp{}
is also equipped with a straightforward program logic that acts as our
``off-the-shelf'' verifier for \imp{} programs. Formally, this logic
proves partial Hoare triples of the form \hltrip{PHI}{c}{PHI}, and is
parameterized over the underlying assertion language. We write
$\sigma \models \phi$ to denote that a state $\sigma$ satisfies the
assertion $\phi$.

Equipped with these ingredients, it is straightforward to state our
relational verification problem:
\begin{definition}[Relational Safety]
  Given a pair of \imp{} programs, \lstinline|c1| and \lstinline|c2|,
  we say that \lstinline|c1| and \lstinline|c2| are \textit{safe} with
  respect to the relational pre- and postconditions $\phi$ and $\psi$
  if every pair of final states reachable from input states meeting
  $\phi$ is guaranteed to satisfy $\psi$. We denote relational safety
  as:
  \begin{align*}
    \footnotesize
    \models_R \{\Phi\} \lstinline|c1| \otimes \lstinline|c2|
    \{\Psi\}  ~\triangleq~
  \forall \sigma_1, \sigma_2.\;
  \sigma_1 \uplus \sigma_2 \models \Phi  \implies
  \: \forall \sigma_1', \sigma_2'.
    \sigma_1, \lstinline|c1| \Downarrow \sigma_1'
    \land \sigma_2, \lstinline|c2| \Downarrow \sigma_2' \implies
  \: \sigma_1' \uplus \sigma_2' \models \Psi
\end{align*}
\end{definition}
\noindent An essential property of this definition is that both
\lstinline|c1| and \lstinline|c2| operate over \emph{disjoint} state
spaces (hence the use of $\uplus$ to merge states)-- as we shall see,
this property plays a key role in the equations used to align
programs. Following the convention of \citet{benton2004}, we use the
subscripts $1$ and $2$ to disambiguate references to any identifiers
shared between the left- and right-hand programs. Thus, the assertion
$\lstinline|x1| > \lstinline|x2|$ is satisfied by any pair of states
$\sigma$ and $\sigma'$ in which $\sigma$ maps $x$ to a larger number
than $\sigma'$.

While several specialized verification techniques for directly
reasoning about relational safety have been proposed, including, e.g.,
\emph{relational} program logics~\cite{benton2004, sousa2016,
  MHRV+19}, our goal here is to reduce a relational verification
problem about a pair of \imp{} programs to logically equivalent claim
involving a \emph{single} intermediate \imp{} program. This claim can
then be established using the program logic for \imp{} that we already
have in hand. Of course, there are many such programs, some of which
are more amenable to automated verification than others. As
\lstinline|double3| demonstrated, aligning similar control flow paths
(e.g., loops) with each other helps a verifier to exploit similarities
between the paths in order to simplify verification. Our strategy is
to represent intermediate programs in a richer domain which explicitly
\emph{aligns} subcomponents of the original programs. This domain is
equipped with relational variants of the control flow structures of
the original program; intuitively, the relational variants group
together control flow paths of the two programs.

\begin{figure}[t!]
  \footnotesize

   \[
   \hspace{-1.25cm}
   \inferrule*[Right=EIfT]{
     \lstinline!st1, b1 ---> true!\\ \lstinline!st2, b2 ---> true!\\\\
     \lstinline! (st1,st2), r2 ---> (st11,st21)!
   }{
     \lstinline!(st1,st2), ifR <<b1|b2>> then r1 else r2 ---> (st11,st21)!
   }
   \qquad\qquad
   \inferrule*[Right=E-Seq]{
     \lstinline!(st1,st2), r1 ---> (st11,st21)! \\\\
     \lstinline!(st11,st21), r2 ---> (st12,st22)!
   }{
     \lstinline!(st1,st2) r1;; r2 ---> (st12,st22)!
   }
 \]

 \[
   \inferrule*[Right=EIfF]{
     \lstinline!sti,bi ---> false!\\
     \lstinline!(st1,st2), r2 ---> (st11,st21)!
   }{
     \lstinline!(st1,st2), ifR <<b1|b2>> then r1 else r2 ---> (st11,st21)!
   }
   \qquad\qquad
   \inferrule*[right=E-Align]{
     \lstinline!st1,s1 ---> st11! \quad
     \lstinline!st2,s2 ---> st21!
   }{
     \lstinline!(st1,st2), <<s1|s2>> ---> (st11,st21)!
   }
 \]

 \[
   \inferrule*[right=EWhileF]{
     \lstinline!sti,bi ---> false!
   }{
     \lstinline!(st1,st2), whileR <<b1|b2>> r ---> (st1,st2)!
   }
   \quad\quad
   \inferrule*[right=E-WhileT]{
     \lstinline!st1,b1 ---> true! \quad
     \lstinline!st2,b2 ---> true! \quad \\\\
     \lstinline!(st1,st2) r ---> (st11,st21)! \\\\
     \lstinline! (st11,st21), whileR <<b1|b2>> r ---> (st12,st22)!
   }{
     \lstinline!(st1,st2), whileR <<b1|b2>> r ---> (st12,st22) !
   }
 \]

 \caption{Big-step semantics of \coreRel{}}
 \label{fig:Semantics}
 \vspace{-1em}
\end{figure}

\paragraph{Syntax} The syntax of aligned programs in \coreRel{} is
given on the righthand side of \autoref{fig:Syntax}. A basic
alignment, \lstinline!<< c1 | c2 >>!, consists of a pair of \imp{}
programs \lstinline!c1! and \lstinline!c2! whose control flows are
completely independent. In contrast, the relational control flow
operators \lstinline!whileR!, \lstinline!ifR!, and \lstinline!;;!
align the control flows of their subexpressions. The branches of the
relational conditional \lstinline!ifR!, for example, are themselves
aligned programs. \autoref{fig:Syntax} also defines some additional
notations which capture common alignment strategies. \lstinline!<<s|]!
and \lstinline![|s>>! embed a single \imp{} program into the left and
right sides of a relational representation, respectively. We also
write \lstinline|whileSt| to denote `stuttered' versions of aligned
loops. Intuitively, the left and right hand loop bodies execute a
different number of times at each loop iteration. The aligned
expression %
\lstinline!whileSt 2 1 <<b1 | b2>> <<c1 | c2>>!, for example,
represents a loop that executes \lstinline|c1| twice for every
execution of \lstinline|c2|.

\paragraph{Semantics} The big-step operational semantics of \coreRel{}
are given by the rules shown in \autoref{fig:Semantics}. As aligned
programs are used to encode the behaviors of two programs, the
reduction relation%
\lstinline!(st1, st2) r ---> (st11, st21)! states that the aligned
command \lstinline!r! takes a pair of disjoint initial states to a
pair of disjoint output states. The basic alignment
\lstinline!<< c1 | c2 >>! yields a pair of output states by combining
the results of independently executing \lstinline!c1! and
\lstinline!c2! (\textsc{E-Align}). Evaluating the aligned program
\lstinline!<<x := 2 | y := 3>>!, for example, results in a pair of
final states where the value of \lstinline|x| in the first state is
\lstinline|2| and the value of \lstinline|y| in the second state is
\lstinline|3|; the value of \lstinline|y| (\lstinline|x|) is left
unchanged in the first (second) state. Evaluating a relational loop
\lstinline!whileR <<b1 | b2>> <<c1 | c2>>!, in contrast, simulates a
``lockstep'' execution of a pair of loops by executing \lstinline!c1!
and \lstinline!c2! in tandem (\textsc{E-While$_R$T}) until one of its
conditions is falsified (\textsc{E-While$_R$F}).

\paragraph{Embedding and Reification}
Any pair of \imp{} programs \lstinline|c1| and \lstinline|c2| can be
embedded into \coreRel{} as the aligned program %
\lstinline!<<c1 | c2>>!. Importantly, this embedding preserves the
semantics of the original pair of programs:
\begin{theorem}[Embedding is Sound]
  \label{thm:embedding+sound}
  The pair of \imp{} programs, \lstinline!c1! and \lstinline!c2!, is
  semantically equivalent to their embedding in \coreRel{},
  \lstinline!<<c1 | c2>>!:
  \begin{align*}
    \forall \lstinline!st1 st2 st11 st21!.~~ &
    \lstinline!st1, c1 --> st11! ~~\land~~   \lstinline!st2, c2 --> st21! ~~\implies~~
    \lstinline!(st1, st2) <<c1 | c2>> ---> (st11, st21)!
  \end{align*}
\end{theorem}

\begin{wrapfigure}{r}{.41\linewidth}
  \vspace{-.5cm}
  \begin{tabular}{L@{\hspace{2pt}}L@{\hspace{6pt}}}
    \lstinline![[<<s1 | s2>>]] :== [[s1]]L; [[s2]]R! \\
    \lstinline![[r1 ;; r2]] :== [[r1]]; [r2]]! \\
    \lstinline![[whileR <<b1 | b2>> r ]] :== !\\
    \quad \lstinline!while ([[b1]]L /\ [[b2]]R) [[r]]! \\
    \lstinline![[ifR <<b1 | b2>> then r1 else r2]] :== ! \\
    \quad \lstinline!if ([[b1]]L /\ [[b2]]R) then [[r1]] else [r2]]! \\
  \end{tabular}
  \vspace{-.25cm}
  \caption{Reification of \coreRel{} into \imp{}}
  \label{fig:DenoteImp}
  \vspace{-.25cm}
\end{wrapfigure}
More importantly, every \coreRel{} alignment can be \emph{reified}
back into a \imp{} program via the $\llbracket\cdot\rrbracket$
function shown in \autoref{fig:DenoteImp}. This function uses a pair
of renaming functions, $\llbracket\cdot\rrbracket_L$ and
$\llbracket\cdot\rrbracket_R$, to ensure that variables that come from
the lefthand side of the aligned program are distinct from those on
the right. The control flow of a reified program mimics that of the
aligned program that produced it. Each iteration of the reified
version of an aligned loop %
\lstinline![[whileR <<b1 | b2>> <<c1 | c2>>]]! simulates lockstep
evaluation of \lstinline![[c1]]! and \lstinline![[c2]]!, for example,
while reifying a pair of unaligned loops
\lstinline![[<<while b1 c1 | while b2 c2>>]]! produces a program that
fully evaluates \lstinline![[while b1 c1]]! before
\lstinline![[while b2 c2]]!, obscuring any intermediate relationships
beween variables and \lstinline![[c1]]! and \lstinline![[c2]]!.

On their own, the reification and embedding functions enable us to
reduce a relational safety problem about a pair of programs
\lstinline!c1!  and \lstinline!c2! to one involving a single program,
i.e., \lstinline![[<<c1| c2>>]]!. When coupled with a notion of
equivalence on aligned programs, however, they provide the formal
foundation for defining a space of intermediate programs that are
sufficient for relational reasoning.

\begin{definition}[Alignment Equivalence]
  Two aligned programs are equivalent if they take the same pair of
  initial states to the same pair of final states:
\begin{align*}
  \lstinline!r1! \equiv \lstinline!r2!
  \triangleq
  \forall \lstinline!st1 st2 st11 st21!.\;
    \lstinline!(st1, st2) r1 ---> (st11, st21)!
    \Leftrightarrow
    \lstinline!(st1, st2) r2 ---> (st11, st21)!
\end{align*}
\end{definition}

Reification is equivalence preserving, in that reifying equivalent
aligned programs yields equivalent \imp{} programs:
\begin{theorem}[Reification preserves Equivalence]
  \label{thm:reify+preserves+eqv}
  Any equivalent pair of aligned programs \lstinline!r1! and
  \lstinline!r2! represent  equivalent intermediate programs, \lstinline![[r1]]!
  and \lstinline![[r2]]!: \\
  $ \lstinline!r1 == r2!  \: \implies \: \forall \sigma\, \sigma'.\;
  \sigma, \lstinline|[[r1]]| \Downarrow \sigma' \Leftrightarrow
  \sigma, \lstinline|[[r2]]| \Downarrow \sigma'$
\end{theorem}
\noindent
A direct consequence of Theorems \ref{thm:embedding+sound} and
\ref{thm:reify+preserves+eqv} is that we can reduce the relational
verification problem to reasoning about an equivalent intermediate program:
\begin{corollary}
  Given a pair of \imp{} programs, \lstinline!c1! and \lstinline!c2!,
  in order to prove that \lstinline!c1! and \lstinline!c2! are safe
  with respect to a pair of relational pre- and postconditions $\Phi$
  and $\Psi$, it suffices to prove that an equivalent intermediate program
  \lstinline|r| is safe:
  $ \lstinline!<<c1|c2>> == r! ~\land~%
  \vdash \{\Phi\} ~\lstinline|[[r]]|~ \{\Psi\} \implies %
  \models_R \{\Phi\}~\lstinline|c1| \otimes \lstinline|c2|~
  \{\Psi\}$
\end{corollary}

Unfortunately, this corollary does not provide any guidance on which
\lstinline!r! to use. While equivalent aligned programs are
\emph{extensionally} equal, they may be \emph{intensionally}
different, in the sense that one may be more amenable to verification
than the other. We now turn to the problem of how to automatically
construct a good alignment.

\section{The Alignment Algorithm}
\label{sec:alignment}

\label{sec:mcmc-extraction}
\begin{wrapfigure}{r}{0.49\textwidth}
  \vspace{-.5cm}
\begin{minipage}{0.49\textwidth}
\begin{algorithm}[H]
  \DontPrintSemicolon
  \Params{$p_1$ and $p_2$: programs, \\
          \texttt{Cost}: cost metric for alignments, \\
          $\mu$: number of SA iterations}
  \Output{intermediate program $p_1 \times p_2$}
  $\Sigma \leftarrow$ \RandomStates{$\langle p_1 | p_2 \rangle$} \;
  $E \leftarrow$ \CreateEGraph{} \;
  \Insert{$E, \langle p_1|p_2 \rangle$} \;
  \EQSat{$E$, \coreRel{}} \;
  $best \leftarrow$ \ExtractLocal{E} \;
  $Traces \leftarrow$ \Evaluate{\Reify{best}, $\Sigma$} \;
  $\hat{\eta} \leftarrow$ \Cost{best, Traces} \;
  \For{$k \leftarrow 0$ to $\mu$}{
    $\tau \leftarrow$ \Temperature{$k, \mu$} \;
    $N \leftarrow$ \Neighbor{$E$, best} \;
    $Traces \leftarrow$ \Evaluate{\Reify{N}, $\Sigma$} \;
    $\eta \leftarrow$ \Cost{N, Traces} \;
    \If{$\eta < \hat{\eta}\; \lor$  \Jump{$\tau$, best, $\hat{\eta}$, $N$, $\eta$}} {
      $(best, \hat{\eta}) \leftarrow (N, \eta)$ \;
    }
  }
  \Return{\Reify{best}}
\caption{\kestrel{}}
\label{alg:kestrel}
\end{algorithm}
\end{minipage}
\vspace{-1em}
\end{wrapfigure}

We first present our high-level algorithm before continuing to
discussion of our realignment rules, data-driven extraction, and the
details of our implementation. Our approach to constructing an aligned
intermediate program is given in~\autoref{alg:kestrel}. The algorithm
takes as input two programs $\textsf{p}_1$ and $\textsf{p}_2$, a
\texttt{Cost} function over candidate alignments, and a parameter
$\mu$ bounding the number of candidates our data-driven extraction
phase should consider. The programs $\textsf{p}_1$ and $\textsf{p}_2$
are assumed to have disjoint variable namespaces, which can be easily
accomplished through an automated $\alpha$-renaming pass.

\autoref{alg:kestrel} procedes as follows: Line 1 generates a random
set of initial states over which candidate programs will be evaluated
when collecting and scoring execution traces (see
\autoref{sec:extraction}). Lines 2 --- 3 create a new e-graph from
the initial embedding of the input programs, $\langle p_1 | p_2
\rangle$. Line 4 then applies equality saturation
(\autoref{sec:realignment}) using a collection of \coreRel{}
realignment rules (\autoref{sec:rewrite-rules}) to construct the set
of candidate alignments. Next, an initial term is extracted using a
cost function which minimizes the number of unfused loops in each
e-class (Line 5), execution traces are collected for that term (Line
6), and those traces are then used to help compute that term's
\texttt{Cost} (Line 7). The algorithm then proceeds to the data-driven
extraction phase, which uses a simulated annealing loop (Lines 8--14)
to converge on an optimal alignment. Each iteration of this loop
generates a new alignment candidate by perturbing the selection of
representative nodes for the e-classes in the current alignment (Line
10). Execution traces are collected for the new candidate (Line 11)
and the candidate is scored (Line 12). The candidate is adopted or
discarded according to this score and an annealing schedule (Lines 13
--- 14).

\begin{figure}[t]
  \begin{tabular}{p{.37\linewidth}p{.63\linewidth}}
\small
\begin{tabular}{rll}
  \lstinline!<<c1 | c2>> ==!
  & \lstinline!<<c1|];; [|c2>>!
  & \textsc{rel-def} \\
  \lstinline!<<c1; c2|] ==!
  & \lstinline!<<c1|];; <<c2|]!
  & \textsc{hom-l} \\
  \lstinline![|c1; c2>> ==!
  & \lstinline![|c1>>;; [|c2>>!
  & \textsc{hom-r} \\
\end{tabular}
    &
\small
\begin{tabular}{rll}
  \lstinline!<<while b c |] ==!
  & \lstinline!<<if b then c; while b c|]!
  & \textsc{unroll-l} \\
   \lstinline!<<c1|];; [|c2>> ==!
  &  \lstinline![|c2>>;; <<c1|]!
  & \textsc{rel-comm} \\
  \lstinline!r1;; (r2;; r3) ==!
  & \lstinline!(r1;; r2);; r3!
  & \textsc{rel-assoc}
\end{tabular}
  \end{tabular}
\begin{align*}
  \small
  \lstinline!<<while b1 c1 | while b2 c2>> ! \equiv
  &\quad \begin{array}{@{}l}
           \lstinline!whileSt n m <<b1 | b2>> <<c1 | c2>>;;!\\
           \:\: \lstinline!<< while b1 c1 |];;!
           \lstinline![| while b2 c2>>!
      \end{array}
  & \textsc{while-align} \\
  \begin{array}{@{}l}
    \lstinline!<<if b1 then c1 else c2 !
    \\
    \lstinline!  | if b2 then c3 else c4>>!
  \end{array}
  \equiv
  &\quad
    \begin{array}{@{}l}
      \lstinline!ifR <<b1 | b2>> then <<c1 | c3>>! \\
      \:\: \lstinline!else ifR <<b1 | not b2>> then <<c1 | c4>>!\\
      \:\: \lstinline!else ifR <<not b1 | b2>> then <<c2 | c3>>!\\
      \:\: \lstinline!else <<c2 | c4>>!
      \end{array}
      & \textsc{if-align}
  \\
  \lstinline!whileR <<b1 | b2>>  r! \equiv
  &\quad
    \begin{array}{@{}l}
      \lstinline!ifR <<b1 | b2>> then r else <<skip | skip>>;;!\\
      \:\: \lstinline!whileR <<b1 | b2>> r!
      \end{array}
      & \textsc{unroll-both}
  \\
  \lstinline!<<if b1 then c1 else c2 | c3>> ! \equiv
  &\quad \lstinline!ifR <<b1 | true>> then <<c1 | c3>> else <<c2 | c3>> !
  & \textsc{cond-l}
  \\
  \lstinline!<<c1 | if b1 then c2 else c3>> ! \equiv
  &\quad \lstinline!ifR <<true | b1 >> then <<c1 | c2>> else <<c1 | c3>> !
  & \textsc{cond-r}
\end{align*}
\vspace{-0.4cm}
\caption{Selected \coreRel{} realignment laws}
\label{fig:rewritingRules}
\vspace{-1em}
\end{figure}

\subsection{\coreRel{} Rewrite Rules}
\label{sec:rewrite-rules}

Observing that program equivalence is a congruence relation, we frame
the search for a good intermediate program as rewriting problem in which we
attempt to \emph{realign} the na\"{i}ve embedding of a pair of
programs into a form more amenable for automated
verification. \autoref{fig:rewritingRules} provides several
equivalences that we can use to explore the space of possible
alignments. Our notion of equivalence naturally admits any
equivalences on non-relational \imp{} programs; for example, it is
sound to unroll one iteration of a loop on one side of an aligned term
(\textsc{unroll-l}). More interestingly, the richer structure of
\coreRel{} programs also includes a set of rules that allow us to
realign terms. The first three rules (\textsc{rel-def},
\textsc{hom-l}, and \textsc{hom-r}) allow us to de- and re-compose
subprograms into different alignments, while the \textsc{rel-assoc}
rule reassociates relational sequences of statements, and the
\textsc{rel-comm} rule leverages the fact that the left- and
right-hand programs operate over distinct state spaces to rearrange
two embedded programs. Observe that the alignments on the two sides of
\textsc{rel-comm} reify into different intermediate programs:
\lstinline![[<<c1|];; [|c2>>]] := c1; c2!, while %
\lstinline![|[[c2>>;; <<c1|]]] := c2; c1!. A similar rule over
sequences of \imp{} commands \lstinline|c1; c2 == c2; c1| is obviously
incorrect in the general case, as \lstinline|c1| and \lstinline|c2|
may modify the same variables.

The \textsc{while-align} rule is particularly important, as it merge
two loops together so that their bodies execute in lockstep. Note that
since \lstinline!whileSt!  terminates as soon as either condition is
false, \textsc{while-align} must add trailing ``runoff''
\lstinline!while! loops after the joint loop in order for this
equivalence to hold. In the case that the original loops always have
the same number of iterations, these loops will never execute. A
similar argument explains the \textsc{if-align} rule.

\subsection{Realignment via Equality Saturation}
\label{sec:realignment}

Using \coreRel{} as the underlying language, e-graphs can be used to
compactly represent a (potentially infinite) number of program
realignments. By \autoref{thm:reify+preserves+eqv}, extractions from
an e-graph saturated with sound \coreRel{} rewriting rules like those
given in \autoref{fig:rewritingRules} are semantically equivalent to
the na\"{i}ve alignment \textit{by construction}.

\begin{figure}
\begin{tabular}{ccc}
  \includegraphics[width=0.3\textwidth,angle=0]{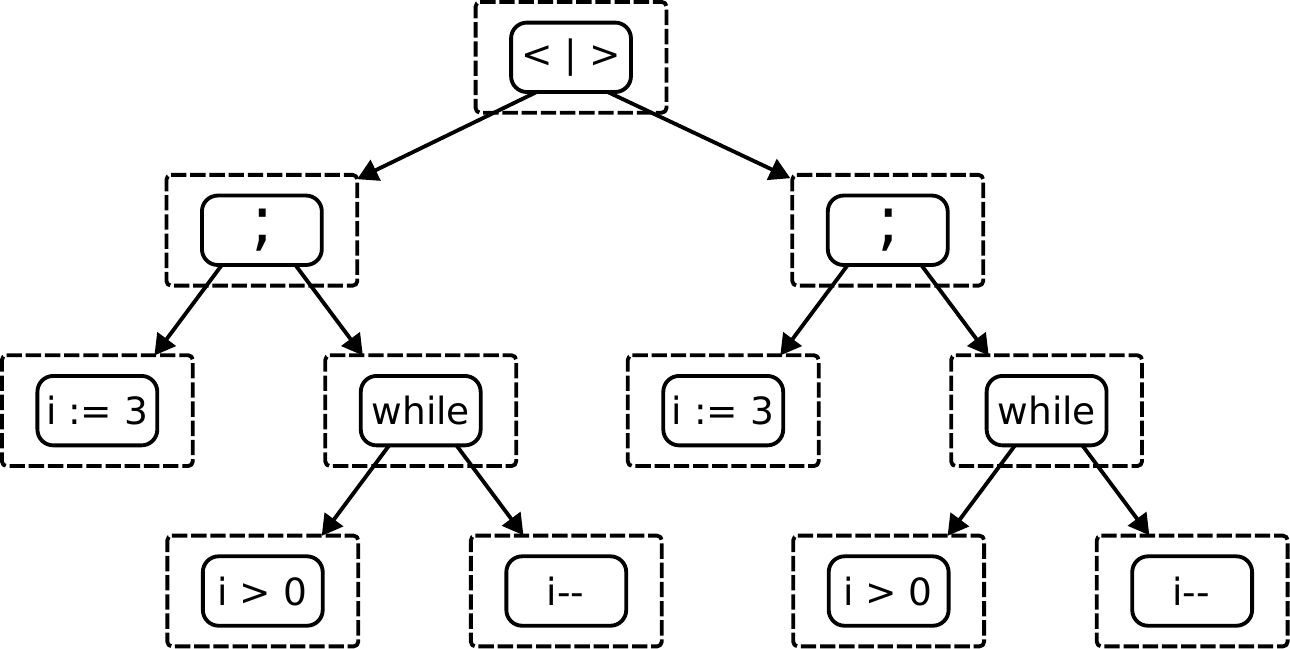}
  & \includegraphics[width=0.3\textwidth,angle=0]{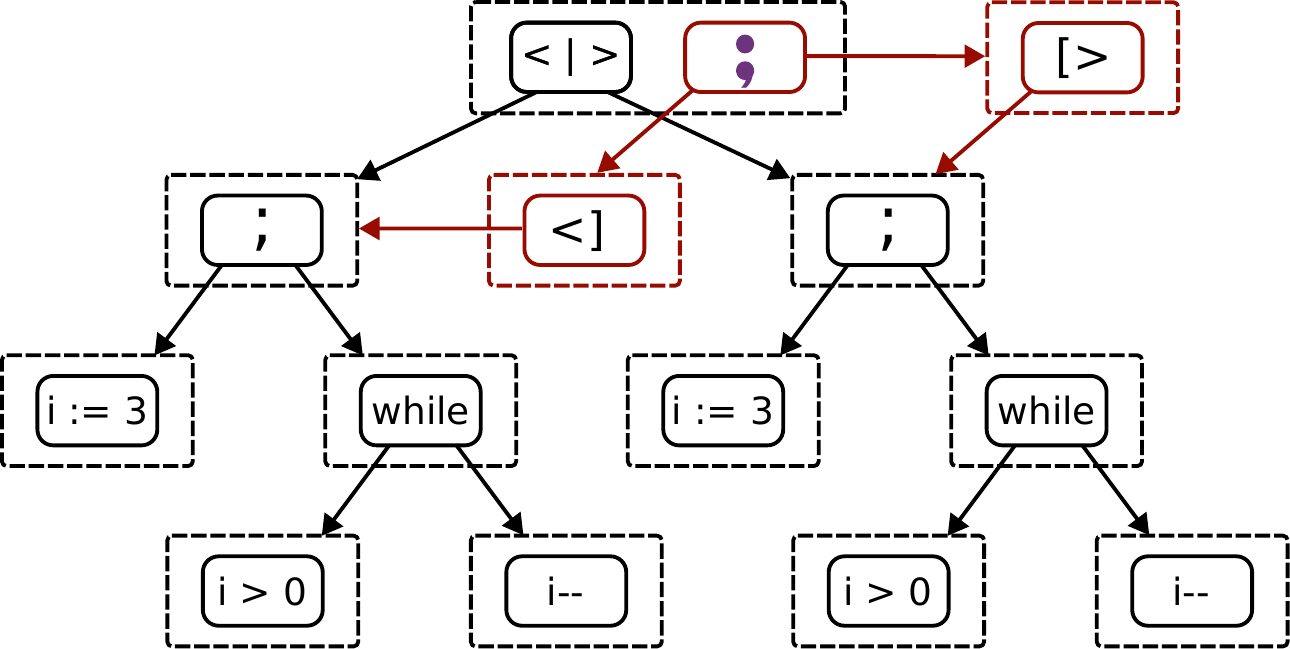}
  & \includegraphics[width=0.3\textwidth,angle=0]{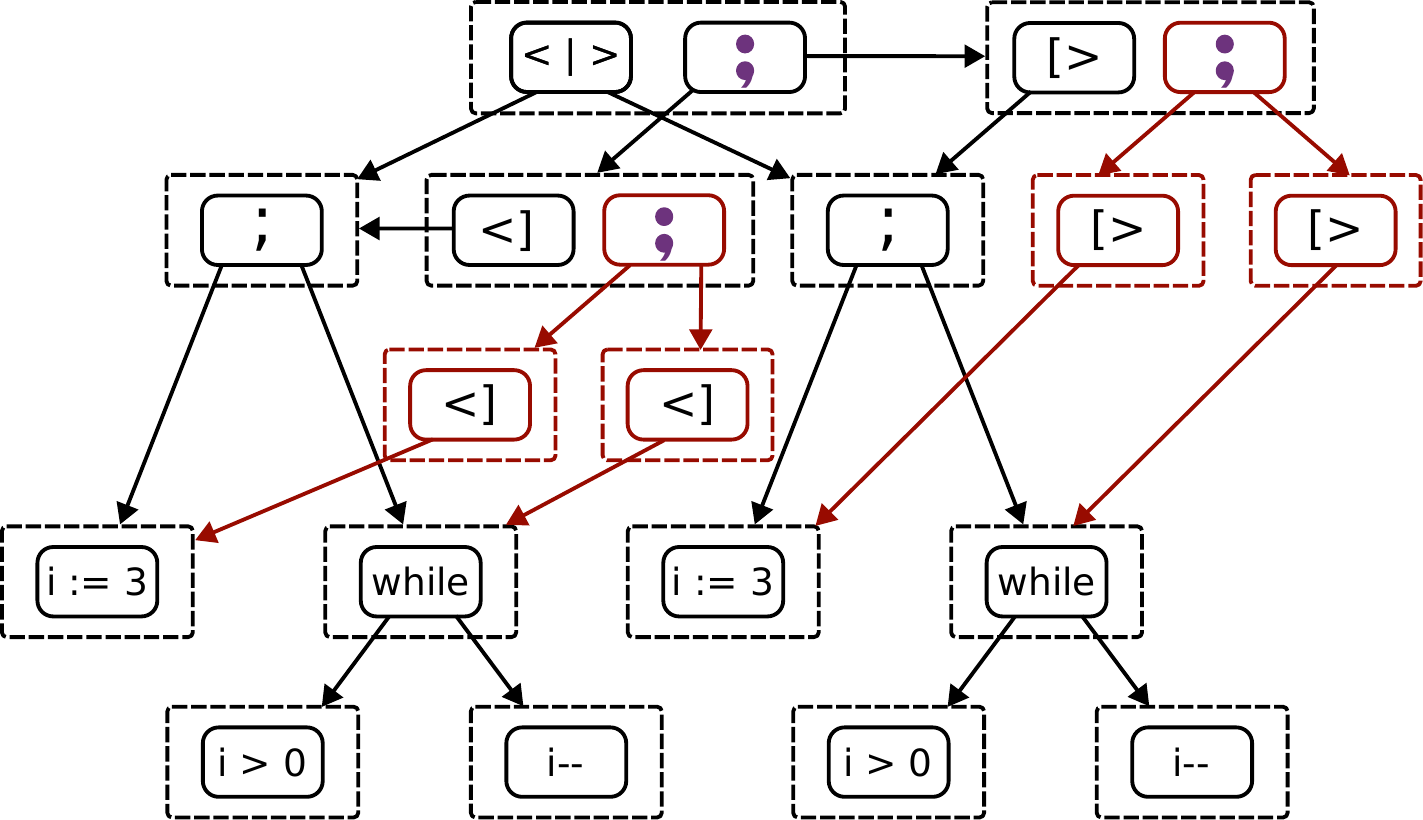} \\
  (a) & (b) & (c)
\end{tabular}
\vspace{-1em}
\caption{E-graphs representing the space of alignments that result
  from applying the rewrite rules from \autoref{fig:rewritingRules} to
  the aligned term %
  \lstinline!<<i := 3; while (i > 0) \{ i--; \} | j := 3; while (j > 0) \{ j --; \}>>!.
  The leftmost e-graph (a) represents this initial alignment. The
  middle e-graph (b) additionally includes the alignment that results
  from applying the \textsc{rel-def} rule. The last e-graph
  (c) includes additional alignments that result from applying both the
  \textsc{hom-l} and \textsc{hom-r} rules. The additional nodes
  that result from each rule are highlighted in
  \textcolor{red}{red}. Some e-nodes
  (\lstinline!<!, \lstinline!:=!, and \lstinline|--|) have been
  combined into a single node for brevity.
}
\label{fig:egraphs}
\vspace{-1em}
\end{figure}

To build an e-graph representing a space of potential alignments of
programs $p_1$ and $p_2$, we start by constructing an e-graph that
contains the na\"{i}ve alignment term \lstinline!<<p1 | p2>>!. For
example, given the \coreRel{} term \lstinline!<<i := 3; !
\lstinline!while (i > 0) { i--; } ! \lstinline!| j := 3; !  \lstinline
!while (j > 0) { j --; }>>!, \autoref{fig:egraphs}(a) depicts the
initial e-graph. We then run equality saturation on the e-graph using
\coreRel{} rewrite rules like those listed in
\autoref{fig:rewritingRules}.

\begin{example}
  \label{ex:eq+sat}
  As an example, the root node of \autoref{fig:egraphs}(a) matches the
  left-hand side of the \textsc{rel-def} rule from
  \autoref{fig:rewritingRules}, with \lstinline|c1| and \lstinline|c2|
  corresponding to the root left and right e-classes,
  respectively. \autoref{fig:egraphs}(b) depicts the e-graph that
  results from applying this rewrite.  Observe that there is a new
  node in the root e-class corresponding to the right-hand side of
  \textsc{rel-def}. We now have a choice when extracting a term from
  this e-graph; if we choose the \lstinline!<< | >>! node in the root
  e-class, we get the original term. If we instead choose the
  \lstinline!;;!  node, we get
  \lstinline!<<i := 3; while (i > 0) { i--; } |]!
  \lstinline!;; [| j := 3; while (j > 0) { j--; }>>!, i.e. the
  original term rewritten according to the \textsc{rel-def} rule.

  Performing additional rewrites to this e-graph will further grow the
  set of equivalent programs it represents.  \autoref{fig:egraphs}(c)
  depicts the e-graph that results from applying \textsc{hom-l} and
  \textsc{hom-r}. Included in the elements of this e-graph is a fully
  decomposed version of the original alignment:
  \lstinline!<<i := 3 |];; <<while (i > 0) { i--; } |]!
  \lstinline!;; [| j := 3>>;; [|while (j > 0) { j--; }>>!.  Further
  applications of the \textsc{rel-comm}, \textsc{while-align}, and
  \textsc{rel-def} rules eventually yield an e-graph that includes the
  (likely) desired alignment: %
  \begin{lstlisting}
<<i := 3; j := 3>>;;
   whileR <<i > 0 | j > 0>> <<i-- | j-- >>;; <<while (i > 0) { i--; } | while (j > 0) { j--; }>>
\end{lstlisting}
\end{example}

\subsection{Data-Driven Extraction}
\label{sec:extraction}

After we have built an e-graph representation of the space of possible
alignments, we still need to \emph{extract} a desirable relational
program which can be reified and handed off to a program
verifier. Before we present our approach to extraction, however, we
first need to define what constitutes a ``good'' alignment. The
ultimate answer is that any alignment that produces a intermediate
program that can be automatically verified is good, and an alignment
that does not is bad. Verification is too expensive to use as a
measure of the quality of a candidate alignment, so we require an
alternative metric. One immediate solution is to define a cost
function that uses syntactic features to identify good alignments.
While such a syntactic approach allows programs to be quickly
extracted, a purely syntactic measure fails to capture important
semantic properties of an alignment. For example, if the ``runoff''
loops generated by an application of \textsc{while-align} never
execute, it suggests a semantic similarity between the loops, as both
loop conditions became false at the same time. However, this semantic
property is not obvious from the syntax of an alignment alone. Our
solution is to combine a syntactic extraction strategy with a
\emph{data-driven} approach~\cite{sharma2013, ZPJ+16, PSM+16, Daikon}
that examines concrete executions of a candidate alignment to
approximate its \emph{semantic} fitness.

\subsubsection{Traces}

The data-driven component of our extraction mechanism executes
candidate alignments in order to gather a set of \emph{traces},
sequences of intermediate states that summarizes the semantic
behaviors of an alignment. The extraction then applies a \emph{cost}
function to each set of traces in order to compare the relative
quality of each alignment. While there are many aspects of an
execution trace which may have bearing on the quality of alignment,
enabling straightforward loop invariants is usually the ultimate goal,
and so we focus on loop executions as the most effective measure of
trace quality. We therefore construct traces of loop head and
end tags:
\begin{enumerate}
\item \whileBegin{}$_R$, \whileHead{}$_R$, and \whileEnd{}$_R$ occur,
  respectively, at the entry, beginning of each iteration, and exit of
  each relational loop (\lstinline!whileR!),
\item \whileBegin{}$_O$, \whileHead{}$_O$, and \whileEnd$_O$ occur in
  similarly for runoff loops generated by \textsc{while-align}, and
\item \whileBegin{}, \whileHead{}, and \whileEnd{} do the same for
  standard \imp{} loops (\lstinline|while|).
\end{enumerate}

\begin{example}
  The program on the left emits the trace on the right when executed
  with a pair of empty initial states:
  \vspace{-1.5em}
  \label{ex:trace1}
  \begin{figure}[H]
    \begin{subfigure}{0.45\textwidth}
      \begin{tabular}{l}
        \lstinline!<<i1 := 3 | i2 := 2>>;;! \\
        \lstinline!whileR <<i1 > 0 | i2 > 0>> <<i1--;| i2--;>>;;! \\
        \lstinline!<<while (i1 > 0) i1++|];;! \\
        \lstinline![|while (i2 > 0) i2-- >>! \\
      \end{tabular}
      \caption{\coreRel{} program.}
    \end{subfigure}
    \hfill
    \begin{subfigure}{0.45\textwidth}
      {\footnotesize
      \begin{align*}
        &\whileBegin{}_R & \textcolor{gray}{\text{-- entering the \lstinline!whileR! loop}} \\
        &\whileHead{}_R, \whileHead{}_R & \textcolor{gray}{\text{-- two iterations of \lstinline!whileR!}} \\
        &\whileEnd{}_R, & \textcolor{gray}{\text{-- exiting \lstinline!whileR!}} \\
        &\whileBegin{}_O, \whileHead{}_O, \whileEnd{}_O& \textcolor{gray}{\text{-- execution of left runoff \lstinline!while! loop}}
      \end{align*}
    }
    \caption{Corresponding trace.}
    \end{subfigure}
    \label{fig:example-trace-1}
  \end{figure}
\end{example}
\vspace{-1em}
\begin{example}
  \label{ex:trace2}
  \lstinline!<<i := 0 | j := 2; while (j > 0) { j--; }>>! emits the
  following trace: {\footnotesize \whileBegin{}, \whileHead{}, \whileHead{}, \whileEnd{}}
\end{example}

\subsubsection{Comparing Alignments}
\label{sec:comparingAlignments}

\begin{figure}[b]
\vspace{-1em}
  \begin{lstlisting}
<<y1 := 0; z1 := 2 * x1 | y2 := 0; z2 := x2>>;;
  whileR <<z1 > 0 | z2 > 0>> <<z1--; y1 := y1 + x1 | z2--; y1 = y1 + x1 >>;;
  <<while (z1 > 0) {z1--; y1 := y1 + x1} | while (z2 > 0) {z2--; y1 = y1 + x1} >>;;
  [|y2 := 2 * y2>>
\end{lstlisting}
  \vspace{-1em}
\caption{A suboptimal alignment of \lstinline|double1| and \lstinline|double2|
from \autoref{sec:overview}.}
\label{fig:suboptimal}
\end{figure}

Before describing our particular cost function over traces, we first
discuss what a desirable trace looks like, using the traces that are
generated by the different alignments of \lstinline!double1! and
\lstinline!double2! from \autoref{sec:overview}.  On one hand, we have
the initial embedding of these programs,
\lstinline!<<double1|double2>>!, and on the other hand we have the
target alignment corresponding to \lstinline!double3!. Consider what
features of the traces generated by \lstinline!double3! indicate that
it should be preferred over \lstinline!<<double1|double2>>!.
An immediate difference is that \lstinline!double3! includes a
relational loop, \lstinline|whileR|, which manifests in its execution
traces as a sequence of \whileHead{}$_R$'s followed by a
\whileEnd{}$_R$. In contrast, the trace of
\lstinline!<<double1|double2>>! contains \emph{only} non-relational
\texttt{wH}'s. This suggests a straightforward heuristic of preferring
traces with more relational loop tags. However, consider the
suboptimal alignment shown in \autoref{fig:suboptimal}. While this
is close to \lstinline|double3| in that it combines loop bodies in a
single \lstinline|whileR|, it does not properly stutter the body of
the relational loop using \lstinline|whileSt|. This manifests in a
trace that includes several \texttt{wH}$_O$'s that are generated by
the lefthand runoff loop after the relational loop has ended
(\texttt{wE}$_R$), suggesting another straightforward strategy of
favoring traces with fewer runoff loop executions.

Based on these observations, we define a cost function for a trace $T$
that has the following two components, where $\#(\texttt{tag}, T)$ is
the number of occurrences of \texttt{tag} in $T$:
\begin{enumerate}
\item The ratio of unmerged to merged loop executions:
\[
  r_{unmerged}(T) = \#(\texttt{wB}, T) / (\#(\texttt{wB}_R, T) + \#(\texttt{wB}, T))
\]
\item The ratio of runoff to non-runoff loop executions:
\[
  r_{runoff}(T) = \#(\texttt{wH}_O, T) / (\#(\texttt{wH}_R, T) + \#(\texttt{wH}_R, T) + \#(\texttt{wH}, T))
\]
\end{enumerate}

The overall cost of a trace is then calculated as $\texttt{cost}(T) =
0.5 * r_{unmerged}(T) + 0.5 * r_{runoff}(T)$.

\subsubsection{Neighboring Alignments}
Our data-driven extraction loop uses the \Neighbor function to
find an alignment ``near'' the current one. Our implementation of
\Neighbor uses an e-graph saturated with \coreRel{} rewrite rules to
locate neighbors as follows:
\begin{enumerate}
  \item Given a \coreRel{} term, find a set of e-classes and choices
    of e-nodes within those classes that represents the term. If
    cycles exist in the e-graph, the same e-class may appear multiple
    times, potentially with a different chosen e-node each time.
  \item Pick one of the e-classes within this set which has more
    than one e-node.
  \item Select a different e-node from this e-class. Assign children
    to this e-node, re-using children of the previous choice when
    possible and randomly choosing children otherwise.
  \item Construct a new \coreRel{} term from this modified set of
    e-node selections.
\end{enumerate}

This strategy assumes that e-nodes in the same e-class represent
alignments which are sufficiently similar to be considered
``neighbors''. This assumption has some caveats, however, as e-nodes
in the same e-class can represent terms an arbitrary number of
rewrites away from each other. Being able to simultaneously apply
multiple rewrites to produce a new alignment is an advantage of our
e-graph-based representation, but this means we must be careful when
identifying an alignment's neighbors, as including terms that are too
dissimilar can cause the search to jump around chaotically,
drastically slowing down convergence. In practice, this becomes an
issue when loop unrolling and scheduling rules place e-nodes
representing arbitrary numbers of unrollings in the same e-class. To
combat this, our implementation of \Neighbor only considers a fixed
number of unrollings one away from the input term.  Additionally,
randomly constructing children of new selections in cases where the
modified choice does not share a child e-class with the original
choice can lead to arbitrarily large subterms. Instead of building
truly random terms in these cases, \kestrel{} biases towards terms
with small ASTs. In practice, we have found that these heuristics
identify good neighbors for our suite of benchmarks.

\subsection{Implementation}
\label{sec:implementation}

We have implemented a relational verification engine based on
\autoref{alg:kestrel}, called \kestrel{}. \kestrel{} is written in
Rust and uses the Egg library~\cite{willsey2021} to represent spaces
of candidate alignments as e-graphs. Internally, \kestrel{} operates
over a superset of \coreRel{}, but it is equipped with a frontend that
accepts a subset of C (it does not support \texttt{for} loops or
\texttt{struct}s, for example) and backends for outputting
intermediate programs in Dafny and C (the latter is used to target
SeaHorn). Our implementation of \kestrel{} hands off the initial
alignment found by syntactic extraction (Line 5) to a verifier; if
this program successfully verifies, \kestrel{} halts and reports its
success. Otherwise, it proceeds to its data-driven extraction phase,
and the result of verifying the intermediate program produced by this
phase is reported to the user.

\paragraph{Equality Saturation Optimizations}
Basic blocks whose internal realignment cannot impact the
verifiability of the intermediate program are encoded using a distinguished
\texttt{basic-block} structure to which the \textsc{hom-l} and
\textsc{hom-r} rules do not apply. This avoids unnecessarily polluting
the search space with useless permutations of realigned straightline
code.

\begin{wrapfigure}{r}{0.5\textwidth}
  \vspace{-.5cm}
\begin{minipage}{0.5\textwidth}
\begin{algorithm}[H]
  \DontPrintSemicolon
  \Params{$p$: aligned intermediate program}
  \Output{loop invariant annotations}
  $I \leftarrow$ \Daikon{$p$} $\cup$ \Hints{$p$} \;
  $p' \leftarrow$ \Annotate{$p$, $I$} \;
  $X \leftarrow$ \Dafny{$p'$} \;
  \While{$X \neq \emptyset$} {
    $I \leftarrow I \setminus X$ \;
    $p' \leftarrow$ \Annotate{$p$, $I$} \;
    $X \leftarrow$ \Dafny{$p'$} \;
  }
  \Return{$p'$}
\caption{HouDafny}
\label{alg:invariant-inference}
\end{algorithm}
\end{minipage}
\vspace{-2em}
\end{wrapfigure}

\paragraph{Instrumentation}
In order to generate traces during its data-driven extraction phase,
\kestrel{} produces instrumented programs that are augmented with
\lstinline|log| commands that produce traces. To generate traces from
an instrumented program, \kestrel{} randomly generates a set of
starting states which meet the verification problem's relational
precondition using test input generators in the style of property
based testing frameworks~\cite{claessen2011quickcheck}.

\paragraph{Reification}
While the work of finding alignments is carried out in the language of
\coreRel{}, \kestrel{} translates \coreRel{} alignments into
intermediate programs annotated with \texttt{assume} and
\texttt{assert} statements. These intermediate programs can be given
directly to off-the-shelf verifiers. Currently \kestrel{} has backends
for C, targeting SeaHorn, and Dafny.

\paragraph{Invariant Inference}
While \kestrel{} tries to find aligned intermediate programs that have
simple loop invariants, not every verification tool, e.g., Dafny,
implements automated invariant inference.  Thus, we have implemented
an invariant inference algorithm, shown in
\autoref{alg:invariant-inference}, for \kestrel{}'s Dafny
backend. This algorithm implements a Houdini-style~\cite{flanagan2001}
iterative refinement invariant inference procedure. Starting from a
pool of candidate invariants that combines invariants suggested by
Daikon~\cite{Daikon} with a built-in set of candidate invariants
(\Hints), the algorithm iteratively tries to verify the program using
Dafny, collecting and removing non-invariant clauses from the set of
candidates each iteration until a fixpoint is reached. The algorithm
showcases one of the strengths of using intermediate programs for
relational verification, as it leverages an existing, non-relational
verification tool, Daikon.

\section{Evaluation}
\label{sec:evaluation}

\begin{table}[t]
\small
\def\arraystretch{0.85}
\caption{Verification results for different alignment
  strategies. Verification failures and successes are marked with
  \textcolor{red}{\ding{54}} and \textcolor{DarkGreen}{\ding{52}},
  respectively. The total time to construct and verify intermediate
  programs are given in seconds; \timeout indicates that a five minute
  timeout was exceeded.  The upper and bottom tables present the
  results for programs with only basic types and ADTs,
  respectively. Benchmark names are annotated with their source:
  \citet{antonopoulos2023} ($^{\dagger}$), \citet{barthe2011}
  ($^{\star}$), \citet{sousa2016}($^\circ$), \citet{shemer2019}
  ($^\diamond$), \citet{clrs} ($^\square$), and
  \citet{churchill2019}($^\ddagger$). For benchmarks with simple
  types, the {\bf Loops} column indicates the presence of a loop in
  the benchmark. For clients of ADTs, the {\bf ADTs} column lists the
  ADTs used-- all these benchmarks had loops. The {\bf Property}
  column gives the relational property being verified. Results in the
  \textbf{Na\"{i}ve} column are for a concatenative alignment
  strategy, the {\bf \kestrel{} (Syntactic)} column presents results
  when only the syntactic component of Kestrel's extraction mechanism,
  while the {\bf \kestrel{} (Full)} column uses both components.}
\label{fig:evaluation-euf}
\begin{tabular}{l|l|l|l|l|l}
  \hline
  \textbf{Benchmark}     & \textbf{Loops}  & \textbf{Property} & \textbf{Na\"{i}ve}
  & \begin{minipage}{.1\textwidth}\centering\textbf{\kestrel{}} \\
    \textbf{(Syntactic)}\end{minipage}
                         & \begin{minipage}{.1\textwidth}\centering \textbf{\kestrel{}} \\ \textbf{(Full)}\end{minipage}\\
  \hline
  commute                   & \OK          & commutativity     & \fail{6.90}    
  & \yay{4.92}         & \yay{4.88}        \\
  data-alignment$^\dagger$    & \OK          & monotonicity      & \fail{8.09}    
  & \fail{8.46}        & \fail{25.50}      \\
  double-square$^\diamond$    & \OK          & equivalence       & \timeout      
  & \fail{21.32}       & \yay{36.22}          \\
  half-square$^\diamond$      & \OK          & noninterference   & \fail{5.04}   
  & \yay{9.06}         & \yay{12.90}       \\
  payments                  & \OK          & equivalence       & \fail{9.12}    
  & \yay{7.22}         & \yay{7.21}        \\
  shemer$^\diamond$           & \OK          & equivalence       & \timeout      
  & \fail{23.05}         & \yay{46.50}      \\
  simple$^\dagger$            & \OK          & equivalence       & \fail{4.35}    
  & \yay{2.60}         & \yay{2.61}        \\
  strength-reduction$^\star$ & \OK          & equivalence       & \fail{8.10}    
  & \yay{6.29}         & \yay{6.31}        \\
  square-sum$^\diamond$       & \OK          & equivalence       & \fail{11.84}  
  & \yay{4.60}          & \yay{4.60}        \\
  unroll$^\star$             & \OK          & equivalence       & \fail{6.26}    
  & \fail{19.10}       & \yay{21.77}       \\
  \hline\hline
  col-item$^\circ$           & \NoOK        & anticommutativity & \yay{2.56}     
  & \yay{2.63}        & \yay{2.59}          \\
  container$^\circ$          & \NoOK        & anticommutativity & \yay{2.58}     
  & \yay{2.69}        & \yay{2.68}          \\
  file-item$^\circ$          & \NoOK        & anticommutativity & \yay{2.52}     
  & \yay{2.56}         & \yay{2.48}        \\
  match$^\circ$              & \NoOK        & anticommutativity & \yay{2.59}     
  & \yay{2.56}        & \yay{2.53}      \\
  node$^\circ$               & \NoOK        & anticommutativity & \yay{2.45}     
  & \yay{2.56}         & \yay{2.65}        \\
\end{tabular} \\
\bigskip
\begin{tabular}{l|l|l|l|l|l}
  \hline
  \textbf{Benchmark}     & \textbf{ADTs}  & \textbf{Property} & \textbf{Na\"{i}ve}
  & \begin{minipage}{.1\textwidth}\centering\textbf{\kestrel{}} \\
    \textbf{(Syntactic)}\end{minipage}
                         & \begin{minipage}{.1\textwidth}\centering \textbf{\kestrel{}} \\ \textbf{(Full)}\end{minipage}\\
  \hline
  array-insert$^\dagger$   & kvstore        & equivalence       & \fail{13.48}   
  & \yay{18.67}        & \yay{14.47}       \\
  array-int$^\circ$       & kvstore        & anticommutativity & \fail{14.44}  
  &  \yay{17.24}       & \yay{17.14}       \\
  bst-min-search$^\square$ & bst            & monotonicity      & \fail{6.61}    
  & \yay{4.54}       & \yay{4.57}        \\
  bst-sum$^\square$        & bst            & monotonicity      & \fail{7.12}    
  & \yay{6.99}       & \yay{7.09}        \\
  bubble-sort$^\star$     & kvstore        & robustness        & \fail{20.41}  
  &  \yay{19.97}       & \yay{17.18}      \\
  chromosome$^\circ$      & kvstore        & anticommutativity & \fail{14.02}   
  &  \yay{20.14}       & \yay{19.90}       \\
  code-sinking$^\star$    & kvstore        & equivalence       & \fail{11.44}   
  & \yay{11.65}       & \yay{11.49}       \\
  flatten$^\ddagger$                & kvstore        & equivalence      & \fail{132.40}  
  & \yay{153.52}      & \yay{10.07}       \\
  linked-list-ni         & list           & noninterference   & \fail{10.71}  
  & \yay{28.48}       & \yay{28.17}       \\
  list-array-sum$^\square$ & kvstore, list  & equivalence       & \fail{7.04}    
  & \yay{5.10}        & \yay{4.96}        \\
  list-length$^\square$    & list           & equivalence       & \fail{4.34}    
  & \yay{3.35}        & \yay{3.23}        \\
  loop-alignment$^\star$  & kvstore        & equivalence        & \fail{11.50}   
  & \fail{17.57}       & \yay{36.68}       \\
  loop-pipelining$^\star$ & kvstore        & equivalence        & \fail{10.87}   
  & \fail{17.56}       & \yay{56.08}       \\
  loop-tiling$^\dagger$    & kvstore        & equivalence        & \timeout      
  & \fail{15.32}         & \fail{83.76}      \\
  loop-unswitching$^\star$ & kvstore        & equivalence        & \fail{8.35}    
  & \yay{5.86}        & \yay{5.96}      \\
  static-caching$^\star$   & kvstore        & equivalence        & \fail{21.63}  
  &  \fail{2.42}       & \fail{185.22}     \\
\end{tabular}
\vspace{-1em}
\end{table}

Our experimental evaluation investigates five key questions regarding our
approach to building aligned intermediate programs:

\begin{itemize}
\item[{\bf RQ1}] Is our approach \emph{effective}, i.e., does
  \kestrel{} construct intermediate programs that enable verification
  tools to be used for relational reasoning?
\item[{\bf RQ2}] Is our approach \emph{expressive} enough to cover a
  diverse set of programs and relational properties?
\item[{\bf RQ3}] Is our approach \emph{efficient}? Is \kestrel{} able
  to find useful intermediate programs within a reasonable time frame?
\item[{\bf RQ4}] How much does each component of \kestrel{} contribute
  to its effectiveness?
\item[{\bf RQ5}] Is our approach \emph{general}? Can \kestrel{} build
  intermediate programs suitable for multiple backend verifiers?
\end{itemize}

\subsection{Benchmark Construction  (\textbf{RQ2})}

To answer these questions, we evaluate \kestrel{} on a diverse corpus
of benchmarks\footnote{All the benchmarks and results from our
  evaluation are provided in the anonymized supplementary material.}
drawn from the relational verification literature, and that includes
examples from both the intermediate program-based and bespoke
approaches~\cite{antonopoulos2023, barthe2011, churchill2019,
  unno2021, sousa2016}. Our benchmark suite includes clients of a
variety of abstract data types (ADTs), including key value stores,
lists, binary search trees, and 2-3 trees ({\bf RQ2}).  Our evaluation
considers several categories of relational properties ({\bf RQ2}),
including:
\begin{itemize}
\item {\bf Equivalence:} Two programs exhibit equivalent behaviors, for
  example always returning the same value given the same inputs.
\item {\bf Anticommutativity:} Swapping the arguments of a function
  inverts its result: \lstinline|compare(a, b) =| \lstinline|!compare(b, a)|, for
  example.
\item {\bf Monotonicity:} Under certain conditions, one program always
  yield a result greater than (or less than) another.
\item {\bf Noninterference:} An information security property that
  requires observable (``low'') outputs of multiple executions to
  independent of any secret (``high'') inputs.
\end{itemize}

\noindent All benchmarks were run on ArchLinux with an 8 core Intel i7-6700K
4GHz CPU and 16 GB RAM.

\begin{figure}[t]
\begin{tikzpicture}
	\begin{axis}[
    height=.25\textheight,
    ybar stacked,
    xtick=data,
    ylabel={Time (s)},
    ymin=0,
    x tick label style={rotate=45,anchor=east},
    tick label style={font=\scriptsize},
    legend style={font=\footnotesize},
    label style={font=\footnotesize},
    axis y line*=none,
    axis x line*=bottom,
    width=.9\textwidth,
    bar width=2mm,
    area legend,
    xticklabels={loop-pipelining, shemer, loop-alignment, double-square, linked-list-ni, unroll, chromosome, bubblesort, array-int, array-insert, square-sum, half-square, code-sinking, flatten, payments, bst-sum, strength-reduction, loop-unswitching, list-array-sum, commute, bst-min-search, list-length, simple, container, col-item, match, node, file-item},
    symbolic x coords={loop-pipelining, shemer, loop-alignment, double-square, linked-list-ni, unroll, chromosome, bubblesort, array-int, array-insert, square-sum, half-square, code-sinking, flatten, payments, bst-sum, strength-reduction, loop-unswitching, list-array-sum, commute, bst-min-search, list-length, simple, container, col-item, match, node, file-item}
  ]
  \addplot coordinates {(square-sum,0.944) (file-item,0.805) (node,0.721) (match,0.717) (col-item,0.741) (container,0.773) (simple,0.852) (list-length,0.811) (bst-min-search, .860) (commute,1.015) (list-array-sum,1.051) (loop-unswitching,1.476) (strength-reduction,0.892) (bst-sum,1.159) (payments,1.345) (code-sinking,1.221) (half-square,1.036) (chromosome,0.885) (array-insert,1.142) (array-int,0.934) (bubblesort,6.508) (unroll,0.901) (linked-list-ni,0.916) (loop-alignment,1.279) (loop-pipelining,1.970) (shemer,1.142) (flatten,2.837) (double-square,.978)};
  \addplot coordinates {(square-sum,11.821) (file-item,1.725) (node,1.813) (match,1.655) (col-item,1.633) (container,1.651) (simple,1.751) (list-length,2.42) (bst-min-search,3.703) (commute,3.864) (list-array-sum,3.913) (loop-unswitching,4.478) (strength-reduction,5.419) (bst-sum,5.929) (payments,5.860) (code-sinking,10.259) (half-square,11.821) (chromosome,18.996) (array-insert,13.281) (array-int,16.178) (bubblesort,10.648) (unroll,18.161) (linked-list-ni,27.244) (loop-alignment,16.130) (loop-pipelining,15.653) (shemer,21.835) (flatten,7.223) (double-square,19.999)};
  \addplot coordinates {(square-sum,0) (file-item,0) (node,0) (match,0) (col-item,0) (container,0) (simple,0) (list-length,0) (bst-min-search,0) (commute,0) (list-array-sum,0) (loop-unswitching,0) (strength-reduction,0) (bst-sum,0) (payments,0) (code-sinking,0) (half-square,0) (chromosome,0) (array-insert,0) (array-int,0) (bubblesort,0) (unroll,0.020) (linked-list-ni,0) (loop-alignment,7.874) (loop-pipelining,27.964) (shemer,9.759) (flatten, 0) (double-square,1.524)};
  \addplot coordinates {(square-sum,0) (file-item,0) (node,0) (match,0) (col-item,0) (container,0) (simple,0) (list-length,0) (bst-min-search,0) (commute,0) (list-array-sum,0) (loop-unswitching,0) (strength-reduction,0) (bst-sum,0) (payments,0) (code-sinking,0) (half-square,0) (chromosome,0) (array-insert,0) (array-int,0) (bubblesort,0) (unroll,0.865) (linked-list-ni,0) (loop-alignment,1.368) (loop-pipelining,1.685) (shemer,1.006) (flatten, 0) (double-square,0.914)};
  \addplot coordinates {(square-sum,0) (file-item,0) (node,0) (match,0) (col-item,0) (container,0) (simple,0) (list-length,0) (bst-min-search,0) (commute,0) (list-array-sum,0) (loop-unswitching,0) (strength-reduction,0) (bst-sum,0) (payments,0) (code-sinking,0) (half-square,0) (chromosome,0) (array-insert,0) (array-int,0) (bubblesort,0) (unroll,1.815) (linked-list-ni,0) (loop-alignment,10.019) (loop-pipelining,8.800) (shemer,12.749) (flatten, 0) (double-square,12.802)};
  \legend{DaikonSyn, HoudiniSyn, Alignment, DaikonSem, HoudiniSem},
	\end{axis}
      \end{tikzpicture}
      \vspace{-1.5em}
\caption{Breakdown of \kestrel{} runtimes by subtask. ``DaikonSyn''
  refers to generating initial invariant candidates for syntactic
  extraction, ``HoudiniSyn'' refers to elimination of invalid
  invariant candidates for syntactic extractions, and ``Alignment''
  refers to semantic extraction of an intermediate program from the
  e-graph.  ``DaikonSem'' and ``HoudiniSem'' refer to the analogous
  invariant inference tasks over semantic extractions. Subtasks which
  take negligible amounts of time (for example, extracting a purely
  syntactic alignment from the e-graph) are not depicted.}
\label{fig:evaluation-subtasks}
\vspace{-1em}
\end{figure}

\subsection{Quality of Aligned Intermediate Programs (\textbf{RQ1,
    RQ3})}
\label{sec:eval+RQ1}
Our first set of experiments addresses our alignment strategy's
ability to produce aligned intermediate programs that can be
effectively and efficiently verified by an existing, non-relational
program verifier, i.e., \kestrel{}'s Dafny backend (\textbf{RQ1,
  RQ3}). We compare \kestrel{} against two baselines: a
\textbf{Na\"{i}ve} strategy which $\alpha$-renames variables and
concatenates the input programs together (see \autoref{sec:overview}),
loops in a program.\footnote{In practice, the syntactic strategy is
equivalent to only using the first component (\textsf{ExtractLocal};
\autoref{alg:kestrel} line 5) of \kestrel{}'s extraction mechanism.}
The \textbf{Syntactic} strategy will always produce merged loops whose
bodies execute in lockstep, but will not attempt to unroll or schedule
loop executions, or perform any other realignments which require
insight into the runtime behavior of the intermediate program.

The results of these experiments are presented in
\autoref{fig:evaluation-euf}. The experiments are grouped into two
tables: the first is comprised of benchmarks that only require basic
datatypes, while the second consists of benchmarks that use ADTs. The
first set is further subdivided into benchmarks with and without loops
(all of our ADT benchmarks contain loops). As the set of paths through
loop-free programs is finite, we expect alignment to be unnecessary
for verification in these cases. Indeed, Dafny is able to verify
all alignments for all five loop-free benchmarks. Nevertheless
these benchmarks show that the alignments computed by \kestrel{} do
not adversely affect verifiability in cases where alignments are not
strictly necessary.

Dafny failed to verify the na\"{i}ve alignments of the remaining 26
benchmarks, suggesting that combining control flow is beneficial.
For all but three of these benchmarks, Dafny was able to verify the
aligned intermediate programs produced by \kestrel{} ({\bf RQ1}). Of
these verified benchmarks, five did not verify with a syntactic
extraction, suggesting a need for semantic methods. In addition,
verification was reasonably efficient, finishing in under 30 seconds
in most cases ({\bf RQ3}). \autoref{fig:evaluation-subtasks} presents
per-subtask timings for the individual components of the \kestrel{}
pipeline. For most of these benchmarks, invariant inference dominates
the total runtime; see discussion below.

Two of the three benchmarks that our pipeline fails to verify require
the insertion of sophisticated guards inside the loops of the aligned
program, transformations that are not currently supported by
\kestrel{}. The \texttt{data-alignment} benchmark must skip executing
certain loop iterations (for example, when the loop index
\texttt{mod} 3 is zero), and \texttt{loop-tiling} requires the
creation of new inner loops which subdivide iterations at certain tile
sizes. The remaining failure case, \texttt{static-caching}, requires
complex loop invariants; a stronger invariant inference engine could
enable the alignments produced by \kestrel{} to be automatically
verified.

\subsubsection{Invariant Inference}

As described in~\autoref{sec:implementation}, the Dafny backend of
\kestrel{} performs a Houdini-style~\cite{flanagan2001} invariant
inference, drawing from a set of predicates generated by
Daikon~\cite{Daikon} and provided by users. This approach requires
multiple verification attempts per benchmark to locate non-invariant
clauses. Our pipeline was able to verify 16 of these benchmarks using
only the predicates suggested by Daikon. Cases which required
additional hints ranged from invariants with simple equalities
between, e.g., loop indices which Daikon did not discover, invariants
which simply carry forward axiomitizations of ADT methods, or
invariants involving more complex relationships requiring insights
into how these ADT axioms interact. Augmenting Daikon with a more
robust hypothesis space or otherwise using invariant inference methods
which specifically target theories of these ADTs has the potential to
improve the success rate without user hints.

\subsection{Contribution of Each Component (RQ4)}

\begin{wrapfigure}{r}{.55\textwidth}
\vspace{-1.5em}
\small
\def\arraystretch{0.85}
\begin{tabular}{l|p{1.2cm}|p{1.2cm}|p{1.2cm}}
  \textbf{Benchmark}     & \textbf{Syntactic} & \textbf{Semantic} &
                                                                    \textbf{Combined}
                                                                    \\
  \hline
  array-insert           & \yay{16.54}      & \yay{34.76}      & \yay{14.47}   \\
  array-int              & \yay{17,35}      & \yay{34.14}     & \yay{17.14}   \\
  bst-min-search         & \yay{4.78}       & \fail{16.52}      & \yay{4.57}    \\
  bst-sum                & \yay{6.88}       & \yay{18.53}       & \yay{7.09}    \\
  bubble-sort            & \yay{18.70}      & \fail{97.14}     & \yay{17.18}   \\
  chromosome             & \yay{19.93}      & \yay{31.45}      & \yay{19.90}   \\
  code-sinking           & \yay{9.30}       & \yay{12.97}      & \yay{11.49}    \\
  col-item               & \yay{2.63}       & \timeout         & \yay{2.59}    \\
  commute                & \yay{4.91}       & \yay{15.19}       & \yay{4.88}    \\
  container              & \yay{2.73}       & \timeout         & \yay{2.68}    \\
  double-square          & \fail{21.03}     & \yay{23.51}      & \yay{36.22}   \\
  flatten                & \yay{9.22}       & \yay{38.11}      & \yay{10.07}    \\
  file-item              & \yay{2.55}       & \yay{57.15}      & \yay{2.48}    \\
  half-square            & \yay{13.10}      & \yay{10.57}      & \yay{12.90}    \\
  linked-list-ni         & \yay{29.01}      & \fail{35.19}     & \yay{28.17}    \\
  list-array-sum         & \yay{5.12}       & \yay{16.93}       & \yay{4.96}    \\
  list-length            & \yay{3.37}       & \timeout         & \yay{3.23}    \\
  loop-alignment         & \fail{19.41}     & \yay{31.15}     & \yay{36.68}   \\
  loop-pipelining        & \fail{15.65}     & \yay{37.89}      & \yay{56.08}   \\
  loop-unswitching       & \yay{6.03}       & \yay{69.28}     & \yay{5.97}    \\
  match                  & \yay{2.71}       & \yay{57.90}      & \yay{2.53}    \\
  node                   & \yay{2.61}       & \yay{10.49}       & \yay{2.65}    \\
  payments               & \yay{7.36}       & \yay{10.59}     & \yay{7.21}    \\
  shemer                 & \fail{23.21}     & \fail{40.17}     & \yay{46.50}   \\
  simple                 & \yay{2.56}       & \yay{2.63}       & \yay{2.61}    \\
  strength-reduction     & \yay{6.28}       & \yay{5.84}     & \yay{6.31}    \\
  square-sum             & \yay{4.50}       & \yay{12.74}       & \yay{4.60}    \\
  unroll                 & \fail{19.15}     & \yay{2.75}       & \yay{21.77}   \\
\end{tabular}
\vspace{-.5em}
\caption{Results of an ablation study over benchmarks with successful
\kestrel{} verifications. The ``Syntactic'' column lists verification
results using just local extraction. The ``Semantic'' column uses
data-driven extractions starting from a na\"{i}ve concatenation. The
``Combined'' column uses the default \kestrel{} workflow. All times
reported in seconds.}
\label{fig:ablation}
\vspace{-1em}
\end{wrapfigure}

To demonstrate the utility of each of the components in \kestrel{}'s
extraction procedure, we have conducted a pair of ablation studies to
evaluate their individual impact on the quality of aligned
programs. The first experiment considers the effectiveness of
\kestrel{}'s combined approach to program extraction. This experiment
uses the benchmarks in \autoref{fig:evaluation-euf} that contain loops
and which our Dafny backend can automatically verify. We gave each of
these benchmarks to two modified versions of \kestrel{}. The first
performs only local extraction (``Syntactic''), while the second
performs \emph{only} the data driven phase, starting from a na\"{i}ve
concatenative alignment (``Semantic''). \autoref{fig:ablation}
presents the results of this experiment.

In most cases, our syntactic extraction technique, which minimizes the
total number of loops (fewer loops likely means more fused loops) is
sufficient for verification. In some cases (e.g.,
\texttt{bst-min-search}, \texttt{linked-list-ni}), this approach
succeeded where data-driven simulated annealing failed; the likely
cause is a large alignment space which causes the MCMC search to
converge too slowly. In several cases (e.g. \texttt{file-item},
\texttt{match}), the programs produced by both approaches were able to
verify, but simulated annealing took much longer than the syntactic
approach. Taken together, these points indicate that using a purely
syntactic extraction is an effective starting point for \kestrel{}'s
simulated annealing approach. For one of these benchmarks
(\texttt{shemer}), neither alignment strategy identified a verifiable
alignment on its own, demonstrating the benefits of \kestrel{}'s
combined approach.

We next evaluated the effectiveness of the cost function used by
\kestrel{}'s simulated annealing loop by constructing two variants of
\kestrel{}: the first uses a cost function that only considers the
number of fused loops as a proportion of all loops, and the second
only considers the number of runoff loop iterations as a proportion of
all loop iterations. We ran an ablation study which examined
\kestrel{}'s performance over the five benchmarks from
\autoref{fig:evaluation-euf} that required semantic extraction.
\autoref{fig:evaluation-cost-ablation} presents the results from this
experiment: in all cases, the combined cost function generated
verifiable alignments where the individual cost function components
alone failed, suggesting that both components of the cost function are
useful in discovering good alignments.

\subsection{Relational Verification with SeaHorn (RQ5)}

\begin{wrapfigure}{r}{0.5\textwidth + 2\FrameSep + 2\FrameRule\relax}
\vspace{-1em}
\begin{tabular}{l|lll}
\textbf{Benchmark} & \textbf{Runoff} & \textbf{Fusion} & \textbf{Combined} \\
\hline
double-square    & \fail{31.84} & \fail{52.38}  & \yay{44.52} \\
loop-alignment   & \fail{33.22} & \fail{57.58}  & \yay{45.82} \\
loop-pipelining  & \fail{36.89} & \fail{66.89}  & \yay{53.50} \\
shemer           & \fail{55.17} & \fail{58.30}  & \yay{45.24} \\
unroll           & \fail{25.54} & \fail{49.02}  & \yay{21.80} \\
\end{tabular}
\vspace{-.5em}
\caption{Results of cost function ablation study across benchmarks
which required semantic extraction. The {\bf Runoff} column considers
only number of runoff loop iterations, the {\bf Fusion} column
considers only fused loops, and the {\bf Combined} column considers
both. All times reported in seconds.}
\label{fig:evaluation-cost-ablation}
\vspace{-1em}
\end{wrapfigure}

To demonstrate that \kestrel{}'s ability to find good intermediate
programs is not tied to a particular verifier ({\bf RQ5}), we
translated a subset of the benchmarks from
\autoref{fig:evaluation-euf} into array-manipulating C programs and
then verified them using \kestrel{}'s SeaHorn backend.
\autoref{fig:evaluation-seahorn} reports the verification times for
na\"{i}ve aligned programs, alignments produced by just the local cost
function (``Syntactic''), and the alignments produced by \kestrel{}'s
default workflow (``Combined''). Our results are grouped into three
categories, shown at the the top, middle, and bottom of
\autoref{fig:evaluation-seahorn}. The top and middle groups comprise
benchmarks where SeaHorn was able to verify the intermediate programs
produced by \kestrel{}, and the top group includes the cases where
SeaHorn was not able to verify the na\"{i}ve alignment. For two of
these benchmarks, \texttt{schemer} and \texttt{simple}, SeaHorn failed
to verify the program found by syntactic methods, while our
data-driven approach was able to find intermediate programs that
successfully verified. Taken together, these results provide evidence
that our approach can support multiple verification backends ({\bf
  RQ4}).

\begin{wrapfigure}{r}{0.5\textwidth + 2\FrameSep +
    2\FrameRule\relax}
\vspace{-1.5em}
\small
\def\arraystretch{0.85}
\begin{tabular}{l|lll}
  \textbf{Benchmark}
  & \textbf{Na\"{i}ve}
  & \begin{minipage}{.1\textwidth}\centering\textbf{\kestrel{}} \\
    \textbf{(Syntactic)}\end{minipage}
                         & \begin{minipage}{.1\textwidth}\centering
                           \textbf{\kestrel{}} \\
                           \textbf{(Full)}\end{minipage}\\
  \hline 
  double-square      & \fail{0.16} & \fail{0.19}  & \yay{1.81}   \\
  shemer             & \fail{0.19} & \yay{0.18}   & \yay{2.05}   \\
  simple             & \timeout    & \yay{0.28}   & \yay{1.68}   \\
  unroll             & \timeout    & \timeout     & \yay{1.58}   \\
  \hline
  array-insert       & \yay{3.52}  & \yay{14.32}  & \yay{19.56}  \\
  array-int          & \yay{0.17} & \yay{0.18}    & \yay{5.35}   \\
  bubble-sort        & \yay{0.14}  & \yay{0.22}   & \yay{16.26}  \\
  chromosome         & \yay{0.15} & \yay{0.16}    & \yay{3.57}   \\
  col-item           & \yay{0.13} & \yay{0.19}    & \yay{2.51}  \\
  container          & \yay{0.16} & \yay{0.15}    & \yay{3.94}  \\
  file-item          & \yay{0.13} & \yay{0.18}    & \yay{21.63} \\
  half-square        & \yay{0.15}  & \yay{0.19}   & \yay{3.98}   \\
  loop-alignment     & \yay{0.13}  & \yay{0.15}   & \yay{6.12}   \\
  loop-unswitching   & \yay{1.19}  & \yay{1.23}   & \yay{6.25}   \\
  match              & \yay{0.14} & \yay{0.16}    & \yay{40.01} \\
  node               & \yay{0.12} & \yay{0.14}    & \yay{2.09}  \\
  \hline
  code-sinking       & \fail{0.67} & \fail{0.29}  & \fail{19.21} \\
  data-alignment     & \timeout    & \timeout     & \fail{33.84} \\
  loop-pipelining    & \fail{5.32} & \fail{15.82} & \fail{15.83} \\
  loop-tiling        & \timeout    & \timeout     & \timeout     \\
  strength-reduction & \timeout    & \fail{0.20}  & \fail{2.74}  \\
  square-sum         & \fail{0.17} & \fail{0.23}  & \fail{1.22}  \\

\end{tabular}
\vspace{-.5em}
\caption{Results from using \kestrel{}'s SeaHorn backend to verify a
  suite of array benchmarks to a suite of array benchmarks. All times
  reported in seconds.
}
\label{fig:evaluation-seahorn}
\vspace{-1em}
\end{wrapfigure}

Analogous to the previous experiment, the middle group of benchmarks
SeaHorn was able to verify using only the na\"{i}ve alignment includes
the loop-free programs at the bottom of
\autoref{fig:evaluation-euf}. However, it additionally contains
several other benchmarks whose na\"{i}ve alignment Dafny was not able
to verify. We conjecture that this is due to SeaHorn's requirement
that programs only use arrays with static sizes, allowing it use
bounded verification techniques not possible in the presence of
datatypes of arbitrary size. As before, while not unlocking new
verifications, these benchmarks provide evidence that \kestrel{}
``does no harm'': programs that verify before alignment continue to
verify after, in a comparable amount of time (minus the overhead of
finding an alignment).

The last group of benchmarks represents the six cases where SeaHorn
was unable to verify \kestrel{} alignments. As before,
\texttt{data-alignment} and \texttt{loop-tiling} require synthesizing
loop conditions currently beyond \kestrel{}'s scope. In the remaining
cases, the complexity of the required loop invariants appears to put
verification out of reach for SeaHorn. To verify these alignments are
nevertheless valid, we manually verified each of these aligned
programs using VST~\cite{vst2018appel}. For the cases verified with
VST, invariants did not require specification of full functional
correctness.

\subsection{Discussion and Limitations}
\label{sec:limitations}

\paragraph{Data-Dependent Alignments}
The alignments \kestrel{} considers are limited by the rewrite rules
given to it. In some cases, these rewrites are insufficient to express
conditions needed for a verifiable alignment. \textit{Data-dependent
  alignments} examine program state to decide control flow, for
example by performing an unrolled or duplicated loop iteration only
when some condition over program variables is met. In general this may
require synthesizing boolean expressions that fall outside the terms
\kestrel{}'s rewrite rules can generate. In the
\texttt{data-alignment} benchmark, for example, a combined loop which
guards the original loop bodies with expressions not found in the
original programs admits a simple loop invariant. In principle, it may
be possible to automatically generate these sorts of guards by
integrating some form of program synthesis into \kestrel{}'s
rewrite-based alignment strategy; we consider this an interesting
direction for future work.

\paragraph{Neighbor and Cost Function Heuristics}
One advantage of constructing an intermediate program is that doing so
enables \kestrel{} to leverage non-relational verifiers
``off-the-shelf''. In order to support as many backend verifiers as
possible, \kestrel{} operates in a verifier-independent way, relying
solely on its neighbor and cost functions to estimate the quality of
an alignment. As a consequence of this deliberate design choice,
however, \kestrel{} can miss alignments that are inconsistent with the
heuristics baked into these functions. In contrast, approaches which
tightly integrate a verifier into the alignment search~\cite{unno2021,
  farzan2019automated, farzan2019reductions, itzhaky2024} are able to
use information from the verifier to discover or refine candidate
alignments, ensuring they do not overlook potentially useful
alignments. One future direction is to explore how \kestrel{} could
similarly leverage feedback from a backend verifier, e.g., the
counterexamples generated by a failed verification attempt, to direct
or inform its search for an alignment.

\paragraph{Backend Verifiers}
While \kestrel{} inherits the strengths of any backend verifier it
targets, this means it also inherits its limitations. (Lack of
invariant inference in Dafny is why we needed to develop an invariant
inference engine based using Daikon for our Dafny backend, for
example.) Similarly, the predicates used to specify the behaviors of
our ADT operations in EUF were not included in Daikon's hypothesis
space; these predicates comprise several of the user-defined
predicates included in our invariant inference engine. We also had to
add several axioms about the semantics of the uninterpreted function
symbols used. For example, the \texttt{loop-alignment} benchmark
needed to be given the invariant $\textsf{left.b} =
\textsf{store}(\textsf{right.b}, \textsf{right.j},
\textsf{read}(\textsf{right.a}, \textsf{right.j}))$ after
the lists \texttt{left.b} and \texttt{right.b} had both stored
values at equal indices from equal locations in equal lists.
While this is a straightforward consequence of the axioms modeling
the lists, these kinds of predicates are outside of Daikon's
predicate space.

\section{Related Work}
\label{sec:related}

\citet{francez1983} identified two high-level strategies for reasoning
about \emph{product properties}, i.e. what we call relational
properties: ``indirect'' methods, which explicitly construct an
auxiliary \emph{product program} that can be verified using standard
techniques for single programs; and ``direct'' approaches, which adapt
single-program techniques to the relational setting. The product
program construction proposed by Francez was equivalent to our naive
construction; he observed that reasoning over this simple
construction often amounted to proving full functional correctness of
the input programs. In response to this challenge, Francez advocated
for techniques tailored to the relational setting, and proposed two
direct approaches: a relational variant of Floyd's inductive
assertions method~\cite{Floyd1967Flowcharts}, and a relational variant
of Hoare logic~\cite{hoare1969}. Instead of discarding indirect
approaches entirely, this paper presents a technique for building
better intermediate programs, constructing an \emph{aligned} program
that exposes the similarities between the input programs in a way that
single-program verifiers can take advantage of.

\subsection{Building Aligned Intermediate Programs}
\citet{barthe2011} also observed that more sophisticated alignment
strategies can yield programs more amenable to verification using
standard techniques. Building on prior work which employed
\textit{self composition}~\cite{barthe2011secure}--- effectively our
naive construction applied to two copies of the same program--- to
formalize information flow properties, \citet{barthe2011} proposed a
set of inference rules for constructing intermediate programs which
\emph{synchronize} the execution of certain instructions, and shows
how they can be used to manually construct several intermediate
programs that can be automatically verified by Why. Subsequent work
extended construction to cover a broader class of \textit{asymmetric
  properties}, which include refinement~\cite{barthe2013}. Neither of
these works present an algorithm for automatically constructing an
intermediate program.

A key challenge to this approach is that the set of intermediate
programs is potentially infinite, so the set of candidate programs
must thus be constrained in some way to find a good alignment. One
strategy to tame this complexity is to restrict the shape of the input
programs. \textit{Hyperproperties} are a restricted form of relational
properties that involve multiple executions of the same
program~\cite{Hyperproperties}. In some sense the two copies of the
same programs are already in perfect alignment; the construction of
\citet{BGAHKS+14} exploits this fact to build a product program
construction for reasoning about differential privacy, which is a
hyperproperty. In order to validate compiler optimizations,
\citet{ZP+08} provides an algorithm for constructing intermediate
programs for \textit{consonant}, i.e., structurally similar, programs
by syntactically aligning a predetermined set of locations in the
source and target programs. The consonance requirement rules out
programs where locations cannot be perfectly synchronized, which
includes programs pairs whose loops execute for different numbers of
iterations, e.g., the programs in \autoref{fig:stutter+loop}.

Richer alignment strategies broaden the set of candidate programs.
Like \kestrel{}, \citet{churchill2019} also try to leverage concrete
program executions to guide the search for intermediate programs.
Instead of trying to align syntactic program locations, that work uses
an \textit{alignment predicate} to identify ``semantic'' points in
program traces that should be aligned, and then constructs a candidate
program based on that alignment. The resulting intermediate program is
not guaranteed to be semantically equivalent to the input programs, so
after construction, candidate are checked to ensure they capture all
the behaviors of the input programs. When the set of traces fails to
cover all the paths in the original program, or a poor alignment
predicate was chosen, this check can fail. In contrast, \kestrel{}
produces intermediate programs that are guaranteed to be equivalent
\textit{by construction}: \kestrel{} first builds a space of
equivalent programs, and then uses data gathered from executions to
identify the solution with the best alignment. In addition, there are
some alignments which cannot be covered by a finite set of executions,
e.g., when the alignment depends on an unbounded input.  Because
\kestrel{} only relies on data to rank (and not construct) candidate
solutions, it is able to find aligned intermediate programs when this
occurs, allowing it to align the programs in
\autoref{fig:data+dependent+loop}, which the approach of
\citet{churchill2019} cannot directly handle.

\subsection{Direct Approaches}
Since Francez first advocated for direct relational verification
techniques, a diverse range of approaches that adapt single-program
verification methodologies to the relational setting have been
proposed. Given the wide range of direct approaches to relational
verification, we limit our discussion to how some of the most
prominent examples exploit relationships between target programs to
simplify relational reasoning.

\paragraph{Relational Program Logics} One particular popular category
of direct techniques is \textit{relational program logics}, which
extend traditional program logics to the relational
setting~\cite{francez1983, benton2004, Yang+07, BKOZ+13, sousa2016,
  banerjee2016, aguirre2017, dickerson2022, MHRV+19, HHL+24,
  dardinier2024hypra, GZAAS+24}. First introduced by Francez, and more
famously rediscovered by \citet{benton2004}, the judgements of these
logics use relational assertions to reason over multiple programs,
and/or multiple copies of the same program. One feature of many of
these relational logics is that they are not syntax-directed:
Cartesian Hoare Logic (CHL)~\cite{sousa2016}, for example, provides
four rules for reasoning about loops, each corresponding to a
different program alignment. CHL's rule for reasoning about loops that
execute in lockstep is roughly analogous to our \textsc{While-Align}
law, for example. Rules corresponding to closer alignments enable
simpler loop invariants, so finding a good alignment in these logics
is often analogous to identifying which of these rules to apply. One
exception is the relational logic of \citet{banerjee2016}, which uses
a ``biprogram'' syntax to allow users to explicitly express alignments
between multiple programs, enabling users to account for different
interleavings of biprogram control flow when constructing proofs.

Automated verifiers based on these logics typically rely on built-in
heuristics to select which rule to apply~\cite{dardinier2024hypra,
  sousa2016}. \citet{chen2019} use reinforcement learning to find
effective proof strategies for different categories of
programs. Alignments are effectively discovered via learned proof
strategies which dictate the order in which to apply rules of the
relational logic. In contrast, \kestrel{} uses concrete program traces
to identify promising alignments \textit{before} verification. An
interesting direction for future work is to investigate if execution
traces could be used to guide rule selection in verifiers based on
relational logics.

\paragraph{Relational Verification via CHC Solving}

Several works have considered how to adapt verification approaches
based on constrained Horn clause (CHC) solving to the relational
setting~\cite{shemer2019, unno2021, itzhaky2024, deangelis2016}. These
approaches generally operate by encoding the semantics of the target
programs as a set of CHCs, and then augmenting or transforming these
in a clauses to enable verification using standard CHC solvers.  This
process often amounts to identifying how to align subparts of the
programs' semantics (which these works sometimes call ``scheduling'').
While these works share our goal of finding and exploiting
correspondences between program structures to enable tractable
automated verification, they all operate directly over an embedding of
the semantics of the target programs and furthermore do not explicitly
construct an intermediate program that is independent from a
particular verifier.

\citet{deangelis2016} first encode the target programs as CHCs and
then constructing an equisatisfiable set of clauses that combines
pairs of predicates into single predicates. This linearizes the
relationships between variables, making resulting in clauses that can
be handled by standard CHC solvers. Inferring how to pair predicates
is tantamount to discovering alignments between subparts of the
represented programs. \citet{shemer2019} verify k-safety properties by
inferring a self-composition function that interleaves control flow
from multiple executions along with an invariant sufficient to verify
the composite program. This procedure is parameterized over a set of
predicates that are used to construct the self-composition and
invariant pair. The inferred self-composition function constitutes an
alignment for the semantic embedding of the target programs.
\citet{unno2021} make alignment constraints manifest in templated
verification conditions, which are expressed in an extension to
constrained Horn clauses. Finding alignment then becomes a concern of
a CEGIS-based verifier for this extended class of CHCs.
\citet{itzhaky2024} present a technique for reducing
$\forall\exists$-hyperproperties over infinite-state transition
systems to a set of CHCs. Program trace alignments are represented
using uninterpreted predicates whose constraints can be expressed as a
CHC problem.

\paragraph{Language-Theoretic Approaches to Relational Verification}
\citet{farzan2019automated, farzan2019reductions} use infinite tree
automata to represent sets of semantic reductions of multiple program
runs, such that proving a relational property for one of these
reductions is sufficient to establish that property over the original
programs. Proofs are discovered via a counterexample-guided refinement
loop, iteratively strengthening a candidate proof until a covered
reduction is found. Valid reductions are defined via a dependence
relation over how original program statements may be reordered. The
infinite tree automata in this approach are roughly analogous to our
use of e-graphs (both data structures represent a space of program
reductions or alignments, respectively), and the dependence relation
controls the space of reductions in a similar way to how our
rewrite rules control the space of possible alignments. Instead of
reducing a semantic model of program traces to produce a proof,
however, our approach uses syntactic realignment laws to construct an
aligned intermediate program. The trace-based models used by
\citet{farzan2019automated, farzan2019reductions} give rise to a
richer space of possible alignments than \kestrel{}'s syntactic
rewrite rules at the cost of requiring verification techniques
specific to that trace model representation. In contrast, the space of
alignments that \kestrel{} can explore is determined completely by the
realignment rules it uses, a more syntactic approach that enables
\kestrel{} to be agnostic to the backend verifier. This strategy also means
that KestRel cannot find alignments outside the scope of its rewrite
rules, e.g., data-dependent alignments. Extending the set of
alignments \kestrel{} can explore while maintaining its independence from
the backend verifier is an important direction for future work.

\subsection{Extracting Terms from E-Graphs}

Extracting a desirable term with a heuristic cost function is a core
piece of synthesis and optimization techniques based on equality
saturation~\cite{tate2009}. Using local cost functions to greedily
select subterms is a common strategy, and forms the default extraction
mechanism of the popular Egg library~\cite{willsey2021}.
\citet{wang2020} propose an alternative non-local approach based on
mixed-integer linear programming (MILP). Although this approach
requires assigning a single, static cost for each e-graph node. In
contrast, alignment problems are most naturally expressed using
variable node costs that depend on, e.g., sibling extractions.
Although it is possible to set up MILP encodings for alignment
problems, our initial experiments using this technique did not scale
to the majority of the benchmarks in our evaluation.

\section{Conclusion}
\label{sec:conclusion}

One way to reason about relationships between multiple programs is to
construct a single intermediate program that captures their behaviors
and then apply standard program reasoning techniques to that
intermediate program to establish the desired relational property. A
key hurdle to employing this strategy is finding an intermediate
program that exposes semantic similarities between the original
programs in way that can be exploited by a verifier. In this paper, we
have presented an approach to automatically building aligned
intermediate program that are amenable to automated verification. We
first embed the target programs in an algebra that captures how they
should be aligned, insert that term into an e-graph, and use
realignment rules to construct a space of aligned intermediate
programs. We use a novel data-driven technique which examines
execution traces to identify the most promising alignment; we extract
an intermediate program from this alignment and hand it off to an
off-the-shelf verifier. We have implemented this approach in a tool,
called \kestrel{}, which supports both Dafny and SeaHorn as backend
verifiers. We have evaluated \kestrel{} on a diverse suite of
benchmarks taken from the relational verification literature. Our
experiments show that \kestrel{} is capable of discovering alignments
that enable verification to succeed where a na\"{i}ve alignment
strategy would otherwise fail.

\section*{Data-Availablity Statement}

The source code for \kestrel{}, our suite of benchmark programs, and a
Coq formalization of \coreRel{} and its metatheory are available via
Github~\cite{kestrel:github} and Zenodo~\cite{kestrel:zenodo}.

\section*{Acknowledgements}

We would like to thank the EGRAPHS community
(\url{https://egraphs.org}) for providing a workshop venue to present
and discuss an initial iteration of this work. We also thank Pedro
Abreu for his help packaging the \kestrel{} code artifact. This
research was partially supported by the National Science Foundation
under Grant CCF-1755880.

\bibliographystyle{ACM-Reference-Format}
\bibliography{bibliography}

\end{document}